\documentclass[
reprint,
superscriptaddress,
 amsmath,amssymb,
 aps, physrev,
]{revtex4-2}

\usepackage{aas_macros}
\usepackage[colorlinks=true,linkcolor=blue,citecolor=blue,urlcolor=blue]{hyperref}

\usepackage{graphicx}
\usepackage{dcolumn}
\usepackage{bm}
\usepackage{amsmath}

\usepackage{mathptmx} 

\usepackage{xcolor}     
\usepackage[normalem]{ulem} 
\usepackage{textgreek}  

\usepackage{upgreek}    
\usepackage{catchfile}  
\usepackage{tikz}       

\usepackage[all]{hypcap}

\def\fauxschelper#1 #2\relax{%
  \fauxschelphelp#1\relax\relax%
  \if\relax#2\relax\else\ \fauxschelper#2\relax\fi%
}
\def\Hscale{.85}\def\Vscale{.74}\def\Cscale{1.12}
\def\fauxschelphelp#1#2\relax{%
  \ifnum`#1>``\ifnum`#1<`\{\scalebox{\Hscale}[\Vscale]{\uppercase{#1}}\else%
    \scalebox{\Cscale}[1]{#1}\fi\else\scalebox{\Cscale}[1]{#1}\fi%
  \ifx\relax#2\relax\else\fauxschelphelp#2\relax\fi}

\newcommand{\sh}[2][]{%
  \CatchFileEdef{\temp}{"|kpsewhich --var-value #2"}{\endlinechar=-1}%
  \if\relax\detokenize{#1}\relax\temp\else\let#1\temp\fi}

\newcommand{\xxx}[1]{}

\newcommand{\cunits}{cm$^{-2}$}

\newcommand{\kmps}{km s$^{-1}$}

\newcommand{\msun}{$M_{\odot}$}

\newcommand{\as}{$A_S$}
\newcommand{\al}{$A_L$}

\newcommand{\cs}{$\chi^2$}

\newcommand{\nh}{$N_{\rm H}$}

\newcommand{\nhz}{$N_{\rm H,Z}$}

\newcommand{\nhs}{$N_{\rm H,S}$}

\newcommand{\nhgal}{$N_{\rm H}^{\rm gal}$}

\newcommand{\sigl}{$\sigma_L$}

\newcommand{\x}{x-ray}

\newcommand{\fek}{Fe~K}
\newcommand{\feka}{Fe~K$\alpha$}

\newcommand{\fekb}{Fe~K$\beta$}

\newcommand{\chandra}{\textit{Chandra}}

\newcommand{\nustar}{\textit{NuSTAR}}

\newcommand{\nicer}{\textit{NICER}}

\newcommand{\xmm}{\textit{XMM-Newton}}

\newcommand{\xrism}{\textit{XRISM}}

\newcommand{\myt}{{\textsc{mytorus}}}
\newcommand{\mlmyt}{{\textsc{ml\_mytorus}}}
\newcommand{\webmlmyt}{{\textsc{web\_ml\_mytorus}}}

\newcommand{\xspec}{{\textsc{xspec}}}

\newcommand{\fr}{Figure~\ref}

\newcommand{\scr}{Sec.~\ref}
\newcommand{\tr}{Table~\ref}
\newcommand{\exi}{\begin{equation}}
\newcommand{\exo}{\end{equation}}

\newcommand{\ten}[2]{$#1\times 10^{#2}$} 

\def\spose#1{\hbox to 0pt{#1\hss}} 
\def\approxlt{\mathrel{\spose{\lower 3pt\hbox{$\sim$}}
        \raise 2.0pt\hbox{$<$}}}
\def\approxgt{\mathrel{\spose{\lower 3pt\hbox{$\sim$}}
        \raise 2.0pt\hbox{$>$}}}

\definecolor{aqua}{rgb}{0.0, 1.0, 1.0}

\newcommand{\sss}{$\sim$}

\begin{document}


\title{\textbf{AGN \x\ reflection spectroscopy with \mlmyt: Neural posterior \\
    estimation with training on observation-driven parameter grids} 
}%

\author{Ingrid Vanessa Daza-Perilla}
\email{vanessa.daza@unc.edu.ar}
\affiliation{Center for Space Science and Technology, University of Maryland,
Baltimore County, 1000 Hilltop Circle, Baltimore, MD 21250, USA}
\affiliation{Laboratory for X-ray Astrophysics, NASA Goddard Space Flight Center,
Mail Code 662, Greenbelt, MD 20771, USA}
\affiliation{Center for Research and Exploration in Space Science and Technology,
NASA Goddard Space Flight Center, Greenbelt, MD 20771, USA}

\author{Panayiotis Tzanavaris}
\affiliation{Center for Space Science and Technology, University of Maryland,
Baltimore County, 1000 Hilltop Circle, Baltimore, MD 21250, USA}
\affiliation{Laboratory for X-ray Astrophysics, NASA Goddard Space Flight Center,
Mail Code 662, Greenbelt, MD 20771, USA}
\affiliation{Center for Research and Exploration in Space Science and Technology,
NASA Goddard Space Flight Center, Greenbelt, MD 20771, USA}
\affiliation{The American Physical Society, Hauppauge, NY 11788, USA}

\author{V.~Madurga-Favieres}
\affiliation{Center for Space Science and Technology, University of Maryland,
Baltimore County, 1000 Hilltop Circle, Baltimore, MD 21250, USA}
\affiliation{Laboratory for X-ray Astrophysics, NASA Goddard Space Flight Center,
Mail Code 662, Greenbelt, MD 20771, USA}
\affiliation{Center for Research and Exploration in Space Science and Technology,
NASA Goddard Space Flight Center, Greenbelt, MD 20771, USA}
\affiliation{Department of Physics, Tor Vergata University of Rome,
Via della Ricerca Scientifica 1, 00133 Rome, Italy}

\author{M.~Yukita}
\affiliation{Laboratory for X-ray Astrophysics, NASA Goddard Space Flight Center,
Mail Code 662, Greenbelt, MD 20771, USA}
\affiliation{Department of Physics and Astronomy, The Johns Hopkins University,
Baltimore, MD 21218, USA}

\author{A.~Ptak}
\affiliation{Laboratory for X-ray Astrophysics, NASA Goddard Space Flight Center,
Mail Code 662, Greenbelt, MD 20771, USA}
\affiliation{Department of Physics and Astronomy, The Johns Hopkins University,
Baltimore, MD 21218, USA}

\author{T.~Yaqoob}
\affiliation{Center for Space Science and Technology, University of Maryland,
Baltimore County, 1000 Hilltop Circle, Baltimore, MD 21250, USA}
\affiliation{Laboratory for X-ray Astrophysics, NASA Goddard Space Flight Center,
Mail Code 662, Greenbelt, MD 20771, USA}
\affiliation{Center for Research and Exploration in Space Science and Technology,
NASA Goddard Space Flight Center, Greenbelt, MD 20771, USA}


\begin{abstract}
\begin{center}
(Received 12 December 2025; accepted 31 July 2026; published 1 October 2026)
\end{center}
  
X-ray spectroscopy of active galactic nuclei (AGN) reveals key information about the circumnuclear matter geometry. Many AGN are known to show a narrow \feka\ line at 6.4 keV and associated Compton-scattered continua in their \x\ spectra, due to primary continuum scattering in cold, neutral material far from the central supermassive black hole. We present a novel approach that uses Simulation-Based Inference with Neural Posterior Estimation (SBI-NPE) to train a machine learning (ML) model based on \nustar\ spectral fitting results from the literature, employing the physically motivated \myt-decoupled model, which distinguishes line-of-sight from global equivalent neutral hydrogen column density (\nhz\ vs. \nhs). To overcome the limitations of traditional frequentist fitting—such as lack of automation, reproducibility, and computational cost—we implement normalizing flows and autoregressive networks to learn flexible posterior distributions from simulated spectra. From 34 \nustar\ spectral fitting results reported in the literature, we generated 34,000 synthetic spectra for both a uniform and a Gaussian parameter distribution, and show that the latter is more strongly observationally driven. We train our neural network to determine four key \myt\ parameters, \nhz, \nhs, power-law spectral index $\Gamma$, and relative normalization \as. Mutual information analysis identifies optimal spectral regions and justifies the inclusion of redshift, exposure time, and Galactic absorption. The observation-based grid significantly outperforms uniform sampling, achieving predictive accuracy for individual parameters greater than 90\% (\nhs\ and \as), 89\% (\nhz), and 82\% for $\Gamma$ within $\pm1\sigma$. For all four parameters simultaneously, predictive accuracy remains high at  $\sim$70\%.  We make \mlmyt, the trained model, publicly available with a web interface that enables fast and reproducible parameter inference from \nustar\ spectra. Its application to NGC 4388 illustrates the promise of the approach. \mlmyt\ offers a reproducible and accessible alternative for exploring \myt-type models and performs well in a statistical sense, highlighting the potential of SBI methods for future studies of distant \x\ reflection in \x\ spectra of AGN and potentially other compact accreting \x\ sources such as Galactic \x\ binaries.
\end{abstract}

\keywords{black hole physics -- radiation mechanisms: general -- scattering -- galaxies: active -- methods: data analysis}
\maketitle


\section{\label{sec-intro}Introduction}

Actively accreting supermassive black holes (SMBHs, \sss$10^6$-$10^9$~\msun) in galaxies with Active Galactic Nuclei (AGN) give rise to a number of prominent features in the \x\ spectral region. When the primary X-ray power-law continuum, thought to originate in a putative hot corona via inverse Compton up scattering of low-energy accretion disk photons \citep[e.g.,][and references therein]{ghisellini1994}, is incident upon relatively cold, neutral material at thousands of gravitational radii from the central engine, a narrow (full width at half maximum, FWHM $<$ 10\,000 \kmps) \feka\ emission line at a rest energy of 6.4~keV due to \x\ fluorescence can be observed \citep[e.g.,][and references therein]{Matt1996}. This line is accompanied by its associated \fekb\ counterpart and a Compton-scattered (``reflected'') continuum, which are both produced in tandem in this distant material. In this work we do not deal with relativistically broadened \feka\ lines and associated material that may arise via processes in the accretion disk close to the SMBH. In what follows, the terms ``\feka\ emission'' and ``line'' will imply the narrow variant only.

Many studies established the ubiquity of a narrow \feka\ line in both Type 1 and Type 2 Seyferts \citep[e.g.,][]{Sulentic1998, Gelbord2001, Weaver2001, Yaqoob2001}. \chandra\ grating spectroscopy enabled \feka\ line width measurements with a resolution a factor of \sss3 greater than that of CCDs, and several key studies found FWHM values predominantly in the range \sss1000 to several thousand \kmps\ \citep[e.g.,][]{yaqoob2004, nandra2006, shu2010, shu2011}.

The spatial origin of this line is believed to lie between the optical Broad Line Region (BLR; \cite{zoghbi2019}) and the Narrow Line Region, extending out to kiloparsec scales \citep[e.g.,][]{marinucci2013, guainazzi2016, marinucci2017, fabbiano2017}. The highest spectral resolution observations \citep[see, e.g.,][for a prominent example with NGC~4151 \xrism\ observations]{xrism2024} make it increasingly evident that the line structure is complex, likely arising from multiple sites at different radii, with relative contributions varying between sources. Upcoming high-resolution observations with \xrism\ and the \textit{NewAthena}'s X-ray Integral Field Unit \citep[X-IFU,][]{barret2023} will enhance this picture even more.

However, a well-known candidate has historically been the putative molecular dusty torus, which forms an integral part of the AGN unification paradigm \citep{antonucci1993, urry1995}.

Regardless of the actual details of the geometry, it is critical for any physical interpretation of the line and associated continua that any modeling of the observed \x\ spectral features consistently take into account the physical processes which lie at their origin. At the very least, a physically consistent model should produce the line and reflected continuum in tandem, since these are coupled via atomic physics. During the last three decades or so, the modeling of these features has become increasingly more sophisticated, and a number of models that show such physical consistency are now available and have been used extensively \citep{ikeda2009, brightman2011, liu2014, balokovic2018, buchner2019, vandermeulen2024, Furui_2016}.

In this work, we make use of such a prominent model, namely \myt\ \citep{murphy2009, yaqoob2012}, in its ``decoupled'' or ``uncoupled'' configuration. A distinctive feature of the decoupled configuration is the ability to constrain a global equivalent neutral hydrogen column density due to Compton scattering, \nhs, separately from the column density due to absorption of the ``zeroth-order'' direct continuum along the line of sight, \nhz. In practice, such a setup corresponds to a clumpy, non-uniform ``torus'' geometry \citep[see][for further details and illustrative sketches]{yaqoob2012, 2019ApJ...885...62T}.

Until now, the sophistication and complexity of models such as \myt\ have meant that model fitting could only be performed in a highly interactive way, and all literature on AGN (and also Galactic \x\ binaries) that makes use of this model, or others with a similar level of sophistication, discusses work done in this way, usually within a dedicated \x\ fitting environment such as \xspec\ \citep{Arnaud_1996}, by means of frequentist \cs\ or $C$-stat minimization \citep[but see][]{buchner2014}. While a plethora of seminal results have been obtained through this methodology, it has at least two main limitations. First, reproducibility is a key component of the scientific method \citep[e.g.,][]{stodden2014}. While the model and data may be publicly available, the fitting procedure is, to a certain extent, a time-consuming, trial-and-error process whose final result is not necessarily straightforward to reproduce quickly and efficiently. Second, both the number and spectral quality of observational datasets are steadily increasing. As an example, \textit{NewAthena} X-IFU, with its revolutionary microcalorimeter technology, large field of view, and large effective area, will provide high-resolution spectra for each individual pixel in the field. One can imagine a time in the not too distant future when thousands of \x\ spectra will be available, each of a spectral quality that allows fitting such, or even more, complex models, routinely going beyond simple power laws with just a few extra components such as Gaussians for prominent emission lines. In anticipation of such an era, the time is ripe to investigate different approaches featuring a high degree of automation and reproducibility. It has also been suggested that issues with local, rather than global, minima in spectral fitting might be reduced by means of machine learning (ML) methods \citep{Barret_2024}.

ML techniques offer such an opportunity. In recent years, ML has found several applications in the analysis of large datasets across multiple subfields of astrophysics and cosmology \citep[for reviews see, e.g.,][and references therein]{acquaviva2023, webb2025}. The majority of ML applications rely on human-generated previous results that are used to train an algorithm. In the case of spectral fitting, the goal of training is often to make an algorithm capable of identifying which combination of model components has the highest probability of being the correct one, based on previous experience used for training. In X-rays, there have been a handful of studies that developed ML methods to obtain results traditionally achieved through spectral fitting. For instance, ML algorithms trained on multiwavelength datasets were developed to predict AGN equivalent neutral hydrogen column densities \citep{silver2023}, or to identify Compton-thick AGN using random forest classifiers in deep survey fields \citep{zhang2025}. Recent pioneering results and methods are now starting to offer full alternatives to spectral fitting, where a list of model parameters, with associated errors, is the output of applying a trained ML algorithm \citep[e.g.,][]{dupourque2025}. Only a few authors have successfully tackled this specific problem by applying {\it Simulation-Based Inference} (SBI) with {\it Neural Posterior Estimation} (NPE), showing promising results. Thus, \cite{Ichinohe_2018}, \cite{Parker_2022}, and \cite{Tregidga_2024} used Neural Networks (NNs) to infer physical parameters from X-ray spectra of AGN and Black-Hole X-ray Binaries (BHXRBs). \cite{Barret_2024} and \cite{dupourque2025} further demonstrated the power of SBI-NPE in the context of X-ray spectral data from the Neutron Star Interior Composition Explorer \citep[\nicer,][]{gendreau2012} and the \textit{NewAthena} X-IFU, respectively. More recently, \cite{Barret_3_2026} extended this framework by combining auto-encoder-based dimensionality reduction with likelihood-based importance sampling, obtaining posterior distributions statistically indistinguishable from those derived with nested sampling across several X-ray instruments.

While these works have successfully demonstrated the power of SBI-NPE, applications to complex and highly physically motivated models are still lacking, with the possible exception of \cite{dupourque2025}, who make use of the {\tt relxill} \citep{Dauser_2016} relativistic emission model. In this paper, we demonstrate that the SBI-NPE methodology can be successfully used to predict, with a high degree of confidence, the most likely values for key physical parameters of \myt\ decoupled, a complex physically-motivated model that includes narrow \feka\ line emission. We use best-fit results from the literature to construct a large sample of realistic \nustar-quality simulated spectra, and use them to train a NN, which is then applied to a subset of spectra that were not used for the training, showing extremely promising results for the application of this method to other telescopes and instruments, both for AGN and Galactic X-ray binaries.

The structure of the paper is as follows. In \scr{sec-models}, we discuss the basics of our SBI-NPE methodology. In \scr{sec-data}, we detail the observational basis and the construction of our simulated spectra, as well as the basics of the \myt\ spectral fitting model. We present the pre-processing of the spectra and univariate analysis, the NN training, validation, and testing in \scr{sec-meth}, where we also include a dedicated subsection on the level of generalization. We discuss a case study for an individual observed \nustar\ spectrum in \scr{sec-resu}. We conclude with a summary in \scr{sec-summ}.
\section{ML Algorithm}\label{sec-models}

In this work, we use the \myt\ spectral model in its decoupled configuration, which describes \x\ emission, both direct, as well as Compton-scattered at large distances around accreting compact objects. We train a supervised ML algorithm on simulated \nustar\ AGN spectra that cover a range of values for key \myt\ physical parameters, and then apply it to predict the most likely values of these parameters for other simulated \nustar\ AGN spectra that were not used in the training.

If a detailed physical model is not available, one can interpret the observational data using empirical tools, such as diagnostic diagrams or data-driven modeling within an {\it unsupervised} ML framework. However, when the physical behavior of an astronomical source is relatively well understood, it becomes possible to train {\it supervised} ML algorithms to recover physical parameters directly from the data. In this context, if the output corresponds to discrete categories, the task is a {\it classification} problem; if the output represents continuous quantities, it is a {\it regression} problem. Furthermore, when the goal is to estimate the {\it uncertainty} associated with those continuous predictions, one can adopt a probabilistic formulation, often implemented through Bayesian regression or other likelihood-based methods. This is the case in our work, where we apply a supervised, probabilistic approach to infer the physical parameters of the decoupled \myt\ model from \x\ spectra of AGN with known physical parameter values from the literature.

In \scr{sec-sbi-npe}, we detail the architecture of the supervised ML algorithm, focusing on the statistical formulation and the metrics used to evaluate the precision of the physical parameters determined by it.


\subsection{Simulation-based inference methodology with neural posterior estimation (SBI-NPE)}\label{sec-sbi-npe}

The goal of inference is to obtain the posterior distribution $p(\bm{\theta} \mid \bm{x})$, where $\bm{\theta} \equiv [\theta_1, \theta_2, \theta_3, \theta_4]$ represents the physical model parameters, and $\bm{x}$ is the observation vector, i.e., the photon counts as a function of energy for an X-ray spectrum. This posterior corresponds to a conditional joint probability distribution, as the parameters are correlated and conditioned on the data.

In practice, no analytical expression for $p(\bm{\theta} \mid \bm{x})$ is available. Instead, spectra are simulated for different parameter values and compared with the observed data. The process relies on Bayes' theorem:

\begin{equation}
    p(\bm{\theta} \mid \bm{x}) = \frac{p(\bm{x} \mid \bm{\theta})\, p(\bm{\theta})}{p(\bm{x})},
\end{equation}

where $p(\bm{x} \mid \bm{\theta})$ is the likelihood, $p(\bm{\theta})$ the prior, and $p(\bm{x})$ the evidence or normalization constant. In many complex problems, the likelihood is intractable, preventing the direct use of standard Bayesian methods. This motivates the use of SBI, which iteratively adjusts physical or data-driven models until the simulated spectra resemble the observations, thus obtaining a high-probability posterior distribution.

Here, we apply SBI-NPE, which combines \textit{normalizing flows} and autoregressive neural networks to model highly flexible conditional distributions of the physical parameters from simulated spectra. The key idea is that we can transform a simple base distribution into a complex posterior through a series of invertible transformations, allowing us to sample a target distribution for a given observation.

\subsubsection{Normalizing flow framework}

Normalizing flows rely on the \textit{change of variables theorem} for probability densities. Starting from a simple multivariate Gaussian distribution $\bm{z} \sim N(\bm{0},\bm{I})$, we apply an invertible transformation $\bm{\theta} = f(\bm{z})$ to obtain a more representative distribution. The density of the transformed variable is:

\begin{equation}
    p_\theta(\bm{\theta}) = p_z\left(f^{-1}(\bm{\theta})\right) \cdot \left| \det\left( \frac{\partial f^{-1}}{\partial \bm{\theta}} \right) \right|.
\end{equation}

Equivalently, using the change of variables $\bm{z} = f^{-1}(\bm{\theta})$, the log-density becomes:
\begin{equation}
    \log p_\theta(\bm{\theta}) = \log p_z(\bm{z}) - \log \left| \det\left( \frac{\partial f}{\partial z} \right) \right|,
    \label{eq:2}
\end{equation}

where we used the identity
\begin{equation}
\left| \det\left( \frac{\partial f^{-1}}{\partial \theta} \right) \right| = \left| \det\left( \frac{\partial f}{\partial z} \right) \right|^{-1}.
\end{equation}

The Jacobian determinant accounts for how the transformation changes volumes in probability space. The strategy is to design invertible transformations whose Jacobian is computationally tractable.

\subsubsection{Autoregressive modeling with MADE}

Since our parameter vector includes multiple correlated variables conditioned on the observation vector $\bm{x}$, the conditional joint distribution can be factorized as:

\begin{equation}
\small
 p(\bm{\theta} | \bm{x}) = p(\theta_1\vert \bm{x}) \cdot p(\theta_2 \vert \theta_1, \bm{x}) \cdot p(\theta_3 \vert \theta_1,\theta_2,\bm{x}) \cdot p(\theta_4\vert\theta_1,\theta_2,\theta_3, \bm{x}).
\end{equation}

For example, in the context of our work, the parameter vector
${\boldsymbol \theta} \equiv
 \{ N_{\mathrm{H,Z}}, N_{\mathrm{H,S}}, \Gamma, A_S \}$
represents the physical quantities of the decoupled \myt\ model conditioned on the observed X-ray spectrum $\bm{x}$.
Accordingly, the conditional joint posterior can be factorized as:

\begin{equation}
\small
\begin{split}
p(\bm{\theta} \vert \bm{x}) = &
\, p(N_{\mathrm{H,Z}} \vert \bm{x}) \cdot
p(N_{\mathrm{H,S}} \vert N_{\mathrm{H,Z}}, \bm{x}) \\
& \cdot\, p(\Gamma \vert N_{\mathrm{H,Z}}, N_{\mathrm{H,S}}, \bm{x}) \cdot
p(A_S \vert N_{\mathrm{H,Z}}, N_{\mathrm{H,S}}, \Gamma, \bm{x}).
\end{split}
\end{equation}

This autoregressive structure is naturally handled by the \textit{Masked Autoencoder for Distribution Estimation} (MADE; \citep{Germain_2015}), which enforces sequential dependence by applying masks to the network connections. These masks ensure that each parameter $\theta_i$ depends only on the previous variables $\theta_{<i}$ and on the observed data $\bm{x}$.

\subsubsection{Masked Autoregressive Flow architecture}

To integrate MADE into a normalizing flow, we implement a series of affine transformations. Each component of the base variable $z \sim N(0,I)$ is transformed as:

\begin{equation}
   \theta_i = \mu_i(\theta_{<i}, x) + \sigma_i(\theta_{<i}, x) \cdot z_i,
   \label{eq:affine_transform}
\end{equation}
where $\mu_i$ and $\sigma_i = \exp(s_i) > 0$ are the location and scale parameters predicted by the MADE network, with $s_i$ being the raw neural network output and the exponential ensuring positivity.

\begin{figure*}[t!]
    \centering
    \includegraphics[width=1.0\linewidth,
                 trim={0 5cm 0 1.5cm}, clip]{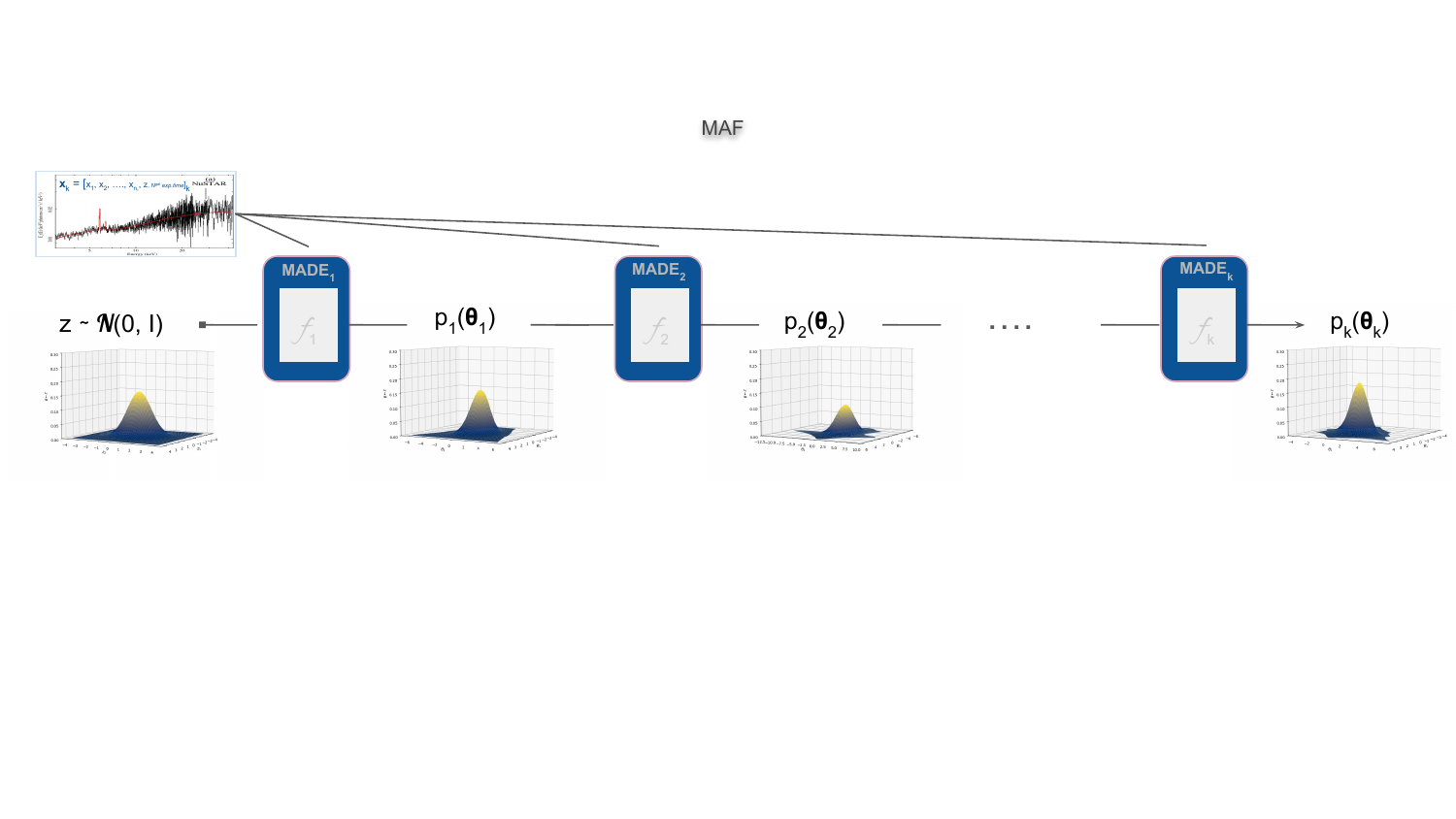}
    \caption{Flow diagram of the SBI-NPE inference mechanism applied to X-ray spectra, implemented via the \textsc{sbi} package. The MAF is illustrated, which transforms latent variables $\bm{z} \sim \mathcal{N}(\bm{0}, \bm{I})$ into progressively more complex posterior distributions considering the information in the spectrum. The 3D plots visualize how the autoregressive dependencies introduced at each step result in increasingly complex distributions, capable of capturing correlations between parameters and non-Gaussian shapes.}
    \label{fig:fig_0}
\end{figure*}

By chaining multiple MADEs into a \textit{Masked Autoregressive Flow} (MAF), we progressively deform the distribution into highly flexible shapes, including multimodal, asymmetric, or heavy-tailed distributions. In our implementation, we use five consecutive MADEs to achieve sufficient expressive power for modeling complex posterior distributions. The scheme of this process is illustrated in Figure~\ref{fig:fig_0}.

\subsubsection{Training procedure}

The neural network is trained with a large set of simulated spectra, four characteristic physical parameters of the observation, as well as $z$, \nhgal, and exposure time. These simulations are obtained with \myt\ in its decoupled mode, described in Section~\ref{sec-data}.

During the training, the network learns to maximize the likelihood of the observed parameter–spectrum pairs by minimizing the \textit{Negative Log-Likelihood} (NLL) loss:

\begin{equation}
    \mathcal{L}_{\rm{NLL}} = - \frac{1}{N} \sum_{k=1}^{N} \log p_w(\theta_k \mid x_k),
    \label{eq:nll_loss}
\end{equation}
where $N$ is the batch size, i.e., the number of examples processed by the network in one training iteration to update the parameters, and $p_w(\theta_k \mid x_k)$ is the posterior probability estimated by the network for parameters $\theta_k$ given spectrum $x_k$, with $w$ denoting the neural network weights.

\subsubsection{Inference and uncertainty quantification}

Once trained, the SBI-NPE model generates posterior samples by sampling from the base distribution $\bm{z} \sim N(0,I)$ and applying the learned transformation $\bm{\theta} = f(\bm{z}, \bm{x})$. To obtain a posterior distribution of $M$ samples for each parameter, the model replicates the input spectrum $M$ times, samples $M$ vectors from the base distribution, and evaluates the transformation.

Since the resulting posterior distributions may be non-Gaussian and asymmetric, we take the mode of each marginal distribution as the point estimator. The associated uncertainty is quantified using $\sigma_{68}$:

\begin{equation}
    \sigma_{68} \equiv \frac{P[84] - P[16]}{2},
\end{equation}
where $P[x]$ denotes the $x$-th percentile of the distribution. This metric captures the central dispersion and is robust against outliers, providing a reliable measure of parameter uncertainty.

The complete operational mechanism and NPE architecture during training and inference stages are described and illustrated in Appendix~\ref{app:C}. The success of this method critically depends on the choice of the prior distribution used to generate the training set, as detailed in Section~\ref{sec-data}.


\section{Prior}\label{sec-data}

In general, studies implementing supervised ML algorithms for \x\ astronomical spectroscopy rely on simulated spectra without considering any observational evidence. For instance, this is the case in \cite{Barret_2024}, where the supervised model is trained on a broad prior that is not representative of any particular observational data. Instead, the authors sample from a broad uniform distribution of values, without any specific observational information.

That approach requires generating a larger training sample or developing strategies such as constraining the prior space according to the spectral characteristics, performing a coarse inference (i.e., a preliminary inference using a small number of simulations to identify the most informative regions of the parameter space and redefine the prior), or, in some cases, repeating this process with several rounds of inference. Although these methods can improve performance on simulated data or on single-test-case observations, they are computationally expensive and may ultimately limit the model’s applicability to several real observations. As an alternative, one can use a different feature space, designed to extract the most relevant spectral information. This is naturally achieved when using observation-based grids. In our approach, we restrict the physical parameter space by directly using information from spectral fits to observational data. We thus build a supervised model that inherits and reproduces the precision inherent in observational spectral fits. We show explicitly (\scr{sec-meth}) that this approach is superior to using a training set where parameter values are chosen from a uniform distribution, which is not observationally based.

At the same time, it is important to caution that, since an observation-driven parameter grid is constructed from literature best-fits obtained with frequentist interactive fitting procedures, an ML algorithm trained on such results may reduce optimizer-related failures during inference but will not necessarily remove biases embedded in the seed fits such as calibration choices, model-misspecification, or local-minimum solutions. A posterior will be conditioned on this literature-anchored prior and may only be able to generalize on new observational data whose inherent, true physical parameters fall within the parameter ranges used for training.

In \scr{sec-mytorus}, we briefly review the \myt\ model in its decoupled mode and its associated physical parameters. In \scr{sec-obs}, we discuss the parameter values compiled from fits in the literature and used to build the training sets. In \scr{sec-sim}, we describe how the simulated spectra, used to train the SBI-NPE model, are generated from the literature-based spectral fitting results. In \scr{sec:samples}, we describe the training, validation, and test samples implemented in the study.

\subsection{\label{sec-mytorus}\myt\ decoupled overview}

We present below a brief overview of \myt\ in its decoupled configuration.
For more detailed descriptions, see \cite{murphy2009}, \cite{yaqoob2011}, \cite{yaqoob2012}, \cite{lamassa2014}, \cite{yaqoob2016}, \cite{tzanavaris2019}, and the \myt\ manual\footnote{\href{http://mytorus.com/manual}{http://mytorus.com/manual}}.

In the \myt-decoupled physical picture, there are three main components.

First, along the line of sight the intrinsic power-law continuum ({\small\tt zpowerlw}  component in \xspec, with power-law index $\Gamma$) undergoes absorption and removal of photons due to Compton scattering. This is modeled with the multiplicative \mlmyt\ \xspec\ table {\small\tt etable\{mytorus\_Ezero\_v00.fits\}}. Thus the total observed, direct, line-of-sight (``zeroth" order, Z) continuum is {\small\tt zpowerlw*etable\{mytorus\_Ezero\_v00.fits\}}. A key physical component associated with the zeroth order table is the equivalent neutral hydrogen column density \nhz\ along the line of sight.

Second, Compton scattering of the intrinsic power-law continuum globally, out of the line of sight, gives rise to a scattered (S, ``reflected") continuum due to the global matter distribution. This is implemented via the additive \myt\ \xspec\ table {\small\tt atable\{mytorus\_scatteredH500\_v00.fits\}}. A key physical component of this table is the equivalent neutral hydrogen column density \nhs, which corresponds to scattering out of the line of sight.

A third additive table ({\small\tt atable\{mytl\_V000HLZnEp000\_v01.fits\}}) implements 
\feka\ and \fekb\ fluorescent line (L) emission, due to neutral iron absorption of energetic \x\ photons that remove K-shell electrons. As for the second table, the column density \nhs\ is also associated with this table because the lines and Compton-scattered continuum are produced in the same global matter distribution.

The normalizations, power-law photon indices ($\Gamma$), and redshift ($z$) values for the three tables are tied, and so are the equivalent neutral hydrogen column densities due to global scattering (\nhs) between the scattered continuum power-law and the \fek\ emission line components.

We should note that the distinguishing characteristic of the decoupled configuration is that there are two distinct equivalent hydrogen column densities, one along the line of sight (\nhz) and one globally out of the line of sight (\nhs). This is unlike the coupled configuration where  where the two column densities are tied together, but the meaning of \nhs\ in coupled mode is that it is the equatorial column density of a torus, so in that case the actual column densities along different lines of sight are strictly determined by the inclination angle of the torus.

In addition, parameters \as\ and \al\ are relative normalization factors of the scattered continuum and fluorescent line emission respectively, with defaults of unity. These parameters allow adjustments that empirically compensate for time lags between the intrinsic continuum and reprocessed emission, and/or compensate for deviations in the data from model assumptions such as covering factor, and solar abundances. In the most common usage of \myt, \al\ is tied to \as, while \as\ is allowed to be a free parameter. \as\ and \al\ are implemented through the {\tt constant} \xspec\ component. 

The implementation of \fek\ emission includes a nominal Gaussian smoothing (\sigl) via an \xspec\ model called {\tt gsmooth}, which is used to model velocity broadening by applying Gaussian convolution to the intrinsic \fek\ line profile. This component is not always implemented in the literature; in this work, we fix it to 100~\kmps, implying a narrow, unresolved line, which is usually the case with \nustar.

Finally, Galactic absorption (\nhgal) is modeled with a {\small\tt phabs} component, which multiplies all of the above setup.

Overall, in \xspec\ command form, the \myt\ decoupled model used here is thus:

\noindent{\small\tt phabs *\\
  (zpowerlw*etable{mytorus\_Ezero\_v00.fits}\\
  +constant*atable{mytorus\_scatteredH500\_v00.fits}\\
  +constant*gsmooth*atable{mytl\_V000HLZnEp000\_v01.fits})\\} 
\vspace{-8pt}\\
where {\small\tt phabs} models \nhgal, and {\small\tt constant} entries model \as~$=$~\al.


\fr{fig-NHplot} illustrates schematically the geometric setup described above.

\begin{figure*}[!htbp]
    \includegraphics[trim={0 450 60 0},clip,width=1\linewidth]{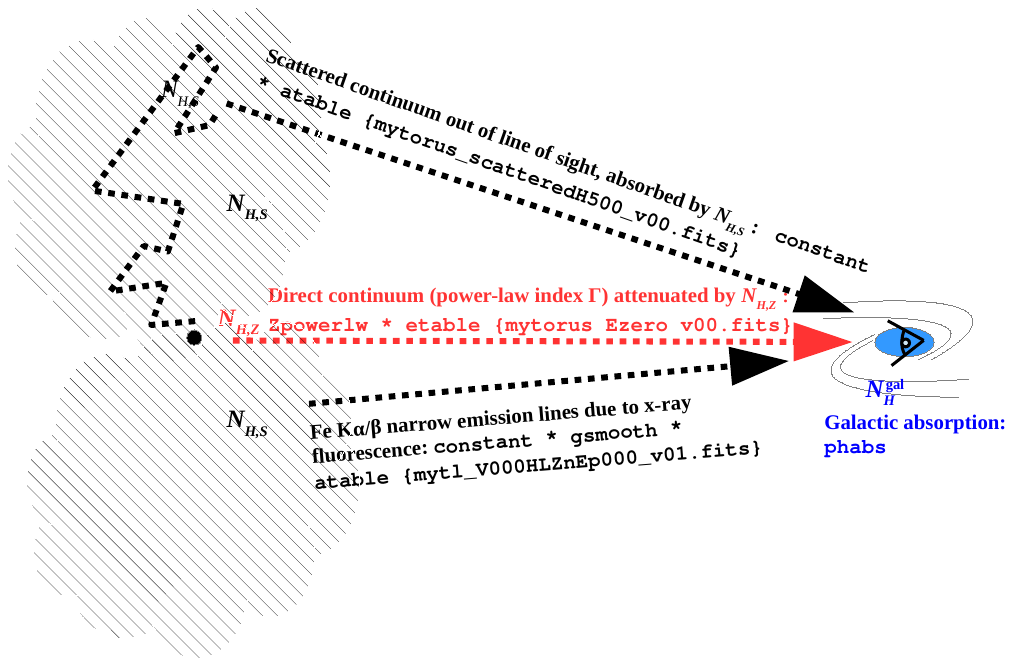}
    \caption{Geometric illustration of the model components for \myt\ decoupled model used in this work. Shown are the direct power-law continuum with power-law index $\Gamma$, associated with a line-of-sight column density \nhz, as well as the scattered continuum and the narrow Fe~K$\alpha$ and $\beta$ emission lines due to \x\ fluorescence from neutral iron both associated with global column density \nhs. Galactic absorption corresponding to a column density \nhgal\ occurs close to the observer. See text for details.}
    \label{fig-NHplot}
\end{figure*}


\subsection{Observations and physical parameters for \myt\ decoupled mode}\label{sec-obs}

In this work, we focus exclusively on observations obtained with \nustar\ \citep{harrison2013}. The motivation for this choice is twofold. First, there is an extensive literature with spectral fitting results for AGN observed with \nustar\ modeled with \myt\ in decoupled mode, which we can use for NN training. In addition, unlike spectra obtained with \chandra\ or \xmm, \nustar\ spectra cover an \x\ spectral range beyond \sss10~keV and into the Compton-hump regime, which is an advantage for models such as \myt. While one can also use contemporaneous observations with different telescopes, this introduces an added degree of complexity in the NN, which we choose to defer for later studies. 

We searched the literature, and compiled fitting results from \cite{torres-alba2023}, \cite{pizzetti2025}, and \cite{jana2022}. These works include results for a sample of 25 AGN in total, mostly Type 2 Seyferts, with the exception of 3C~445, which is Type 1. This sample includes a total of 35 distinct sets of physical parameter measurements and their associated uncertainties. In Table~\ref{tab:tab_1} we show the reported values of the parameters of interest, i.e.~the four main physical parameters of \myt\ in decoupled mode as well as the double exposure time (which corresponds to the combined exposure of the Focal Plane Modules, FPMA and FPMB). Doubling the exposure time is necessary because the Auxiliary Response File (ARF) and Redistribution Matrix File (RMF) used for the simulated data correspond to a single detector. Since the response differences between FPMA and FPMB are negligible because they are contemporaneous, the total effective exposure can be treated as twice that of a single module. We also include \nhgal \footnote{\url{https://heasarc.gsfc.nasa.gov}} and $z$. 

\begin{figure*}[!htbp]
    \centering
    \includegraphics[width=1\linewidth]{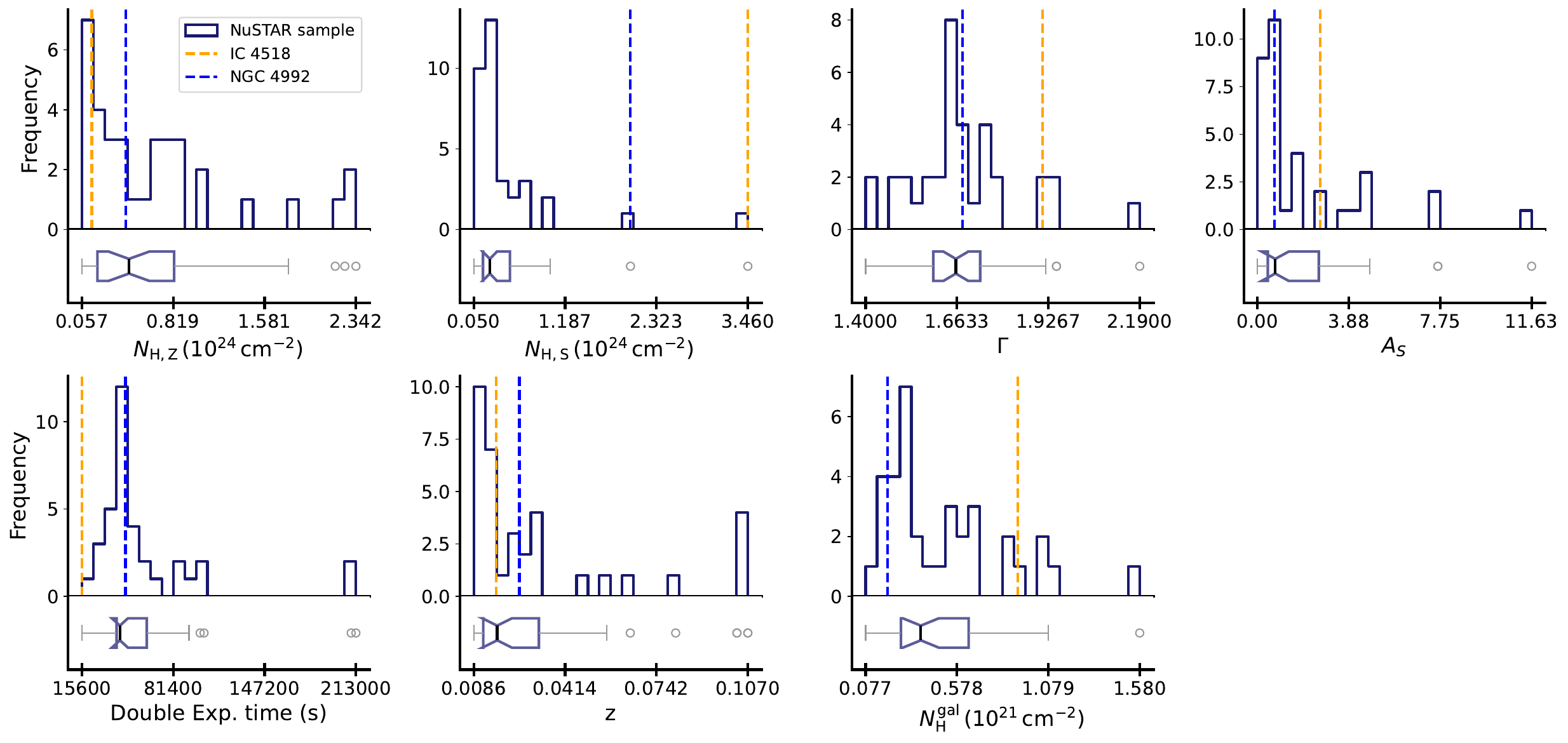}
    \caption{Distributions of literature-based physical parameters for 35 \nustar\ observations (25 AGN). From left to right and top to bottom, shown are distributions for four parameters from \myt\ decoupled fits (\nhz, \nhs, $\Gamma$, \as), as well as (double) exposure time, $z$, and \nhgal. Each panel shows a histogram with a corresponding box-plot below it. 
    }
    \label{fig:fig_1}
\end{figure*}

In Figure~\ref{fig:fig_1} we show the distribution of these parameters by means of histograms and box plots \footnote{In the boxplot diagrams, the right and left vertical line segments of each box represent the 25th and 75th percentiles, while the shorter line segments inside a box represent the medians. Notches indicate the 95\% confidence interval (CI) symmetrically about the median. For skewed distributions or small-sized samples, the CI might be wider than the 25th or 75th percentile, therefore the plot will display a somewhat ``inside out'' shape. The lines extending beyond the boxes are called whiskers. The leftmost whisker extends down to the minimum value where the 25th percentile ends, and the rightmost whisker extends up to the maximum value where the 75th percentile begins. The whisker lengths might not be symmetrical since they must end at an observed data point.}. We also indicate the two AGN IC4518 and NGC4942 with vertical dashed orange and blue lines, respectively. These systems are outliers in \nhs\ space, which in turn makes them outliers in the physical-parameter space of the four key \myt\ parameters used for training and validation. This can also be seen clearly in the (3+1)D plots of \fr{fig:fig_2}. The three spatial axes correspond to the three \myt\ physical parameter values \nhz, \nhs, and $\Gamma$; the parameter \as\ is color coded as indicated by the color bar. Note that data associated with ESO 263-13 are not included, as it was not used in the training because the reported fit had \as~$=0$. The left panel shows the original 34 best-fit values from the literature, while the right panel the 34000 values generated from gaussian distributions centered on the original values. These latter values are the ones used to construct simulated spectra and train the NN algorithm. The two \nhs\ outliers, IC4518 and NGC4942, are marked with orange and blue points in the left panel, corresponding to the dashed lines in \fr{fig:fig_1}. Unlike the case of \as~$=0$, we do not remove these two \nhs\ outliers, since they do not show anomalous behavior for their other \myt\ parameters used in this work (see \fr{fig:fig_1}). Nevertheless, these points affect the precision of SBI-NPE, as shown in \scr{sec-resu}.

\begin{table*}
\caption{\label{tab:tab_1}%
Details of 35 \nustar\ observations and spectral fitting parameters from the literature for the initial sample of 25 AGN. The columns list: source name, double exposure time, \myt\ model best-fit parameters from the literature (two equivalent neutral hydrogen column densities, along the line of sight due to attenuation, \nhz, and globally due to Compton scattering, \nhs; power-law spectral index, $\Gamma$; and relative normalization, \as), followed by \nhgal, $z$, and literature references.
$^{(*)}$ The observation ESO 263-13 was excluded because it corresponds to \as\sss$0$.}
\begin{ruledtabular}
\begin{tabular}{lcccccccc}
    \multicolumn{1}{c}{Source} &  
    \multicolumn{1}{c}{Double} &
    \multicolumn{1}{c}{\nhz} &
    \multicolumn{1}{c}{\nhs} &
    \multicolumn{1}{c}{$\Gamma$} &
    \multicolumn{1}{c}{\as} &
    \multicolumn{1}{c}{\nhgal} &
    \multicolumn{1}{c}{$z$} 
    &\multicolumn{1}{c}{Reference} 
    \\
    
    \multicolumn{1}{c}{} &  
    \multicolumn{1}{c}{Exp. time (s)} &
    \multicolumn{1}{c}{($10^{24}$ \cunits)} &
    \multicolumn{1}{c}{($10^{24}$ \cunits)} &
    \multicolumn{1}{c}{} &
    \multicolumn{1}{c}{} &
    \multicolumn{1}{c}{($10^{21}$ \cunits)} &
    \multicolumn{1}{c}{} 
    &\multicolumn{1}{c}{} 
    \\
    \hline
    NGC 612 & 33400 & 
    0.84 $^{+0.13}_{-0.11}$ & 
    0.67 $^{+1.63}_{-0.33}$ & 
    1.54 $^{+0.16}_{-0.14}$ & 
    0.12 $^{+0.06}_{-0.04}$ & 
    0.172 & 
    0.0299 
    & \cite{torres-alba2023}
    \\
    NGC 788 & 30800 &
    1.10 $^{+0.10}_{-0.09}$ &
    0.19 $^{+0.02}_{-0.02}$ &
    1.92 $^{+0.11}_{-0.12}$ &
    0.92 $^{+0.21}_{-0.16}$ &
    0.215 &
    0.0136  
    & \cite{torres-alba2023}
    \\
    NGC 833 & 41400 & 
    0.18 $^{+0.10}_{-0.10}$  &
    0.06 $^{+0.08}_{-0.05}$  &
    1.69 $^{+0.26}_{-0.25}$  &
    1.00 $^{--}_{--}$ &
    0.227 &
    0.0139 
    & \cite{torres-alba2023} 
    \\
    3C 105 & 41400 & 
    0.45 $^{+0.08}_{-0.07}$ & 
    0.40 $^{+0.57}_{-0.21}$ & 
    1.48 $^{+0.15}_{-0.08}$ & 
    0.75 $^{+0.48}_{-0.40}$ & 
    1.04 & 
    0.1031 
    & \cite{torres-alba2023}
    \\
    3C 105 & 41400 & 
    0.39 $^{+0.04}_{-0.05}$ &
    0.40 $^{+0.21}_{-0.57}$ &
    1.48 $^{+0.15}_{--}$ &
    0.75 $^{+0.40}_{-0.48}$ &
    1.04 &
    0.1031 
    & \cite{torres-alba2023}
    \\
    4C+29.30 & 42000 &
    0.61 $^{+0.17}_{-0.13}$ & 
    0.21 $^{+0.04}_{-0.02}$ & 
    1.72 $^{+0.22}_{-0.20}$ & 
    0.81 $^{+0.19}_{-0.15}$ & 
    0.456 & 
    0.0648 
    & \cite{torres-alba2023}
    \\
    NGC 3281 & 41400 & 
    2.25 $^{+0.24}_{-0.27}$ & 
    0.31 $^{+0.10}_{-0.06}$ & 
    1.65 $^{+0.11}_{-0.12}$ & 
    0.31 $^{+0.30}_{-0.17}$ & 
    0.661 & 
    0.0107 
    & \cite{torres-alba2023}
    \\
    NGC 4388 & 42800 & 
    0.30 $^{+0.01}_{-0.01}$ & 
    0.10 $^{+0.01}_{-0.01}$ & 
    1.58 $^{+0.01}_{-0.01}$ & 
    0.53 $^{+0.12}_{-0.12}$ & 
    0.257 & 
    0.0086 
    & \cite{torres-alba2023}
    \\
    NGC 4388 & 100800 & 
    0.22 $^{+0.004}_{-0.005}$ & 
    0.10 $^{+0.01}_{-0.01}$ & 
    1.58 $^{+0.01}_{-0.01}$ & 
    0.53 $^{+0.12}_{-0.12}$ & 
    0.257 & 
    0.0086 
    & \cite{torres-alba2023}
    \\
    IC 4518 & 15600 & 
    0.14 $^{+0.04}_{-0.03}$ & 
    3.46 $^{+6.54}_{-1.29}$ & 
    1.91 $^{+0.15}_{-0.14}$ & 
    2.65 $^{+0.75}_{-0.58}$ & 
    0.913 & 
    0.0166 
    & \cite{torres-alba2023}
    \\
    3C 445 & 39800& 
    0.33 $^{+0.03}_{-0.03}$ & 
    0.14 $^{+0.02}_{-0.01}$ & 
    1.75 $^{+0.07}_{-0.07}$ & 
    4.26 $^{+7.29}_{-4.26}$ & 
    0.913 & 
    0.0564 
    & \cite{torres-alba2023}
    \\
    NGC 7319 & 29400 & 
    2.17 $^{+0.36}_{- 0.26}$ & 
    0.25 $^{+0.07}_{-0.04}$ & 
    1.73 $^{+0.15}_{-0.17}$ & 
    0.15 $^{+0.27}_{-0.15}$ & 
    0.642 & 
    0.0228 
    & \cite{torres-alba2023}
    \\
    NGC 7319 & 83800 & 
    1.78 $^{+0.34}_{- 0.34}$ & 
    0.25 $^{+0.07}_{-0.04}$ & 
    1.73 $^{+0.15}_{-0.17}$ & 
    0.15 $^{+0.27}_{-0.15}$ & 
    0.642 & 
    0.0228 
    & \cite{torres-alba2023}
    \\
    3C 452 & 103600 & 
    0.39 $^{+0.03}_{- 0.03}$ & 
    0.05 $^{+0.01}_{-0.01}$ & 
    1.53 $^{+0.05}_{-0.05}$ & 
    2.55 $^{+0.46}_{-0.40}$ & 
    0.872 & 
    0.0811 
    & \cite{torres-alba2023}
    \\
    %
    2MASXJ06\footnote{2MASX J06411806+3249313} &
    36600 & 
    0.09 $^{+0.03}_{- 0.02}$ & 
    0.35 $^{+0.16}_{-0.16}$ & 
    1.73 $^{+0.12}_{-0.13}$ & 
    11.63 $^{+10.50 }_{-5.86}$ & 
    1.58 & 
    0.0470 
    & \cite{pizzetti2025}
    \\ 
    NGC 454E & 48400 & 
    1.098 $^{+0.25}_{- 0.23}$ & 
    0.11 $^{+0.02}_{-0.01}$ & 
    1.66 $^{+0.04}_{-0.04}$ & 
    1.55 $^{+0.21}_{-0.19}$ & 
    0.199 &
    0.0120 
    & \cite{pizzetti2025}
    \\
    MRK 348 & 43000 & 
    0.06 $^{+0.0041}_{- 0.0040}$ & 
    0.30 $^{+0.02}_{-0.02}$ & 
    1.61 $^{+0.01}_{-0.01}$ & 
    4.68 $^{+0.48}_{-0.42}$ & 
    0.576 & 
    0.0150 
    & \cite{pizzetti2025}
    \\
    MRK 348 & 209600 & 
    0.07 $^{+0.0034}_{-0.0032}$ & 
    0.30 $^{+0.02}_{-0.02}$ & 
    1.61 $^{+0.01}_{-0.01}$ & 
    4.68 $^{+0.48}_{-0.42}$ & 
    0.576 & 
    0.0150 
    & \cite{pizzetti2025}
    \\
    NGC 4992 & 47000 & 
    0.42 $^{+0.04}_{- 0.04}$ & 
    2.00 $^{+0.35}_{-0.32}$ & 
    1.68 $^{+0.06}_{-0.07}$ & 
    0.71 $^{+0.35}_{-0.37}$ & 
    0.197 & 
    0.0250 
    & \cite{pizzetti2025}
    \\
    ESO 383-18 & 34600 & 
    0.14 $^{+0.02}_{- 0.01}$ & 
    1.00 $^{+0.03}_{-0.05}$ & 
    1.79 $^{+0.03}_{-0.03}$ & 
    7.65 $^{+0.91}_{-1.08}$ & 
    0.380 & 
    0.0120 
    & \cite{pizzetti2025}
    \\
    ESO 383-18 & 213000 & 
    0.15 $^{+0.01}_{- 0.01}$ & 
    1.00 $^{+0.03}_{-0.05}$ & 
    1.79 $^{+0.03}_{-0.03}$ & 
    7.65 $^{+0.91}_{-1.08}$ & 
    0.380 & 
    0.0120 
    & \cite{pizzetti2025}
    \\
    MRK 417 & 41400 & 
    0.30 $^{+0.04}_{- 0.03}$ & 
    0.69 $^{+0.32}_{-0.25}$ & 
    1.64 $^{+0.07}_{-0.09}$ & 
    1.93 $^{+0.87}_{-1.60}$ & 
    0.191 & 
    0.0320 
    & \cite{pizzetti2025}
    \\
    MCG-01-05-047 & 26800 & 
    2.34$^{+0.54}_{- 0.43}$ & 
    0.21 $^{+0.05}_{-0.03}$ & 
    1.66 $^{+0.18}_{-0.15}$ & 
    4.77 $^{+2.22}_{-1.96}$ & 
    0.285 & 
    0.0170 
    & \cite{pizzetti2025}
    \\
    ESO 103-35 & 54600 & 
    0.18 $^{+0.01}_{- 0.01}$ & 
    0.60 $^{+0.29}_{-0.21}$ & 
    1.95 $^{+0.02}_{-0.02}$ & 
    1.67 $^{+0.14}_{-0.12}$ & 
    0.584 & 
    0.0130 
    & \cite{pizzetti2025}
    \\
    ESO 103-35 & 87600 & 
    0.19 $^{+0.01}_{- 0.01}$ & 
    0.60 $^{+0.29}_{-0.21}$ & 
    1.95 $^{+0.02}_{-0.02}$ & 
    1.67 $^{+0.14}_{-0.12}$ & 
    0.584 & 
    0.0130 
    & \cite{pizzetti2025}
    \\

    NGC 1142 & 41400 & 
    1.47 $^{+0.12}_{- 0.13}$ & 
    0.22 $^{+0.02}_{-0.01}$ & 
    1.65 $^{+0.09}_{-0.05}$ & 
    0.15 $^{+0.00}_{-0.15}$ & 
    0.556 & 
    0.0280 
    & \cite{pizzetti2025}
    \\

    IRAS 16288+3929 &
    32200 & 
    0.83 $^{+0.120}_{- 0.102}$ & 
    0.29 $^{+0.06}_{-0.06}$ & 
    2.19 $^{+0.11}_{-0.11}$ & 
    0.11 $^{+0.00}_{-0.11}$ & 
    0.0774 & 
    0.0300 
    & \cite{pizzetti2025}
    \\

    ESO 263-13$^{(*)}$ & 45600 & 
    0.81 $^{+0.06}_{- 0.08}$ & 
    0.08 $^{+0.02}_{-0.02}$ & 
    1.69 $^{+0.05}_{-0.05}$ & 
    0.00 $^{--}_{--}$ & 
    1.08 & 
    0.0320 
    & \cite{pizzetti2025}
    \\

    Fairall 272 & 48600 & 
    0.13 $^{+0.02}_{- 0.02}$ & 
    0.74 $^{+0.37}_{-0.23}$ & 
    1.53 $^{+0.11}_{-0.09}$ & 
    3.49 $^{+1.68}_{-1.33}$ & 
    0.450 & 
    0.0220 
    & \cite{pizzetti2025}
    \\

    LEDA 2816387 & 53000 & 
    0.64 $^{+0.07}_{- 0.07}$ & 
    0.09 $^{+0.14}_{-0.05}$ & 
    1.40 $^{+0.00}_{-0.00}$ & 
    0.34 $^{+0.95}_{-0.4}$ & 
    0.312  & 
    0.1070 
    & \cite{pizzetti2025} 
    \\
    LEDA 2816387 & 49200 & 
    0.75 $^{+0.10}_{- 0.09}$ & 
    0.09 $^{+0.14}_{-0.05}$ & 
    1.40 $^{+0.00}_{-0.00}$ & 
    0.34 $^{+0.95}_{-0.4}$ & 
    0.312 & 
    0.1070 
    & \cite{jana2022}
    \\
    NGC 4507 & 60266 & 
    0.82 $^{+0.13}_{- 0.13}$ & 
    0.22 $^{+0.06}_{-0.05}$ & 
    1.67 $^{+0.04}_{-0.05}$ & 
    0.71 $^{+0.03}_{-0.05}$ & 
    0.312 & 
    0.0118 
    & \cite{jana2022}
    \\
    NGC 4507 & 68928 & 
    0.70 $^{+0.14}_{- 0.13}$ & 
    0.22 $^{+0.06}_{-0.05}$ & 
    1.65 $^{+0.08}_{-0.09}$ & 
    0.70 $^{+0.02}_{-0.08}$ & 
    0.312 & 
    0.0118 
    & \cite{jana2022} 
    \\
    NGC 4507 & 64450 & 
    0.81 $^{+0.15}_{- 0.15}$ & 
    0.22 $^{+0.06}_{-0.05}$ & 
    1.66 $^{+0.06}_{-0.06}$ & 
    0.60 $^{+0.06}_{-0.08}$ & 
    0.312 & 
    0.0118 
    &  \cite{jana2022}
    \\
    NGC 4507 & 61848 & 
    0.71 $^{+0.13}_{- 0.14}$ & 
    0.22 $^{+0.06}_{-0.05}$ & 
    1.64 $^{+0.06}_{-0.09}$ & 
    0.71 $^{+0.05}_{-0.04}$ & 
    0.312 & 
    0.0118 
    & \cite{jana2022}
    \\

\end{tabular}    
\end{ruledtabular}
\end{table*}

\begin{figure*}[!htbp]
    \centering
    \includegraphics[width=0.48\linewidth,
                 trim={0 0 0 1.5cm}, clip]{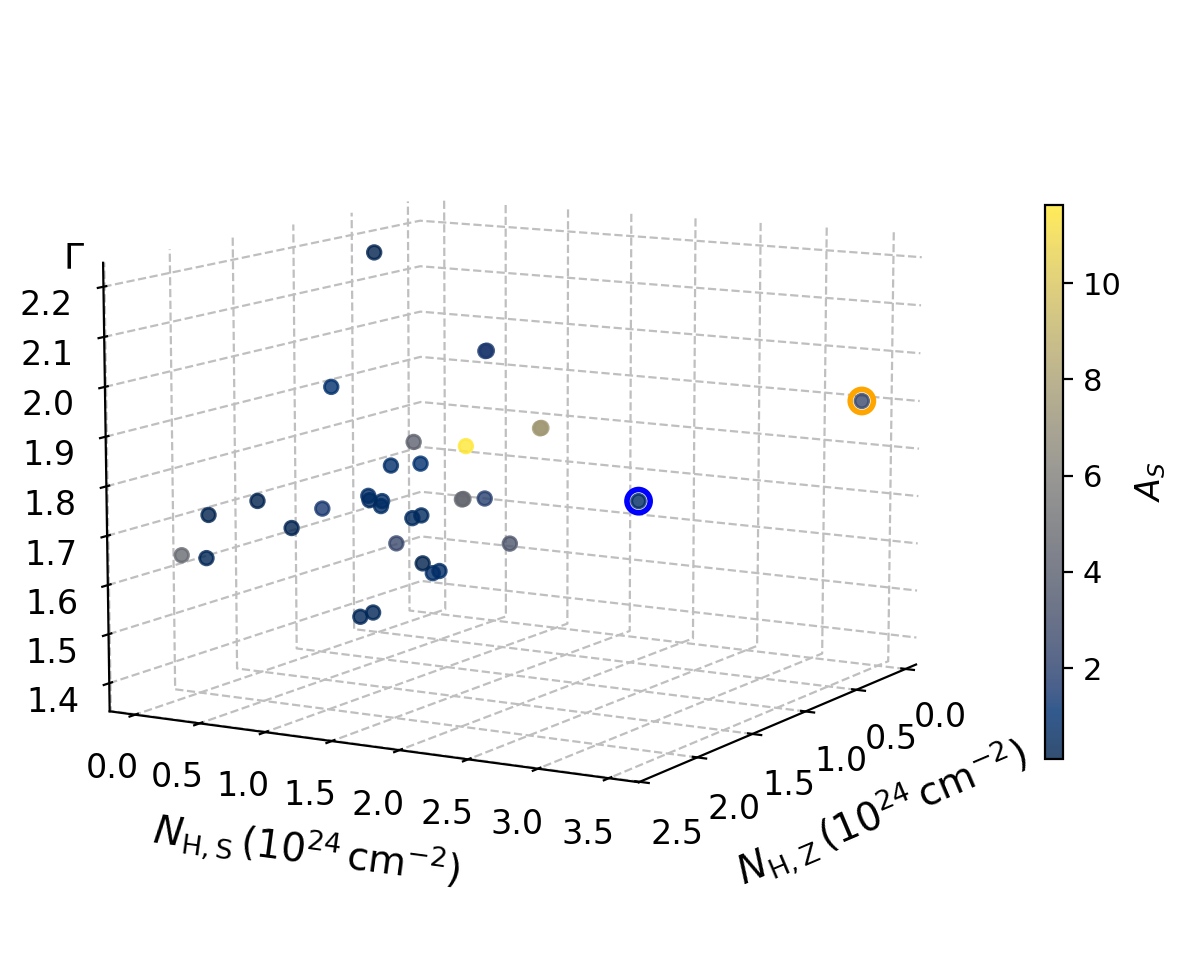}
    \includegraphics[width=0.48\linewidth,
                 trim={0 0 0 1.5cm}, clip]{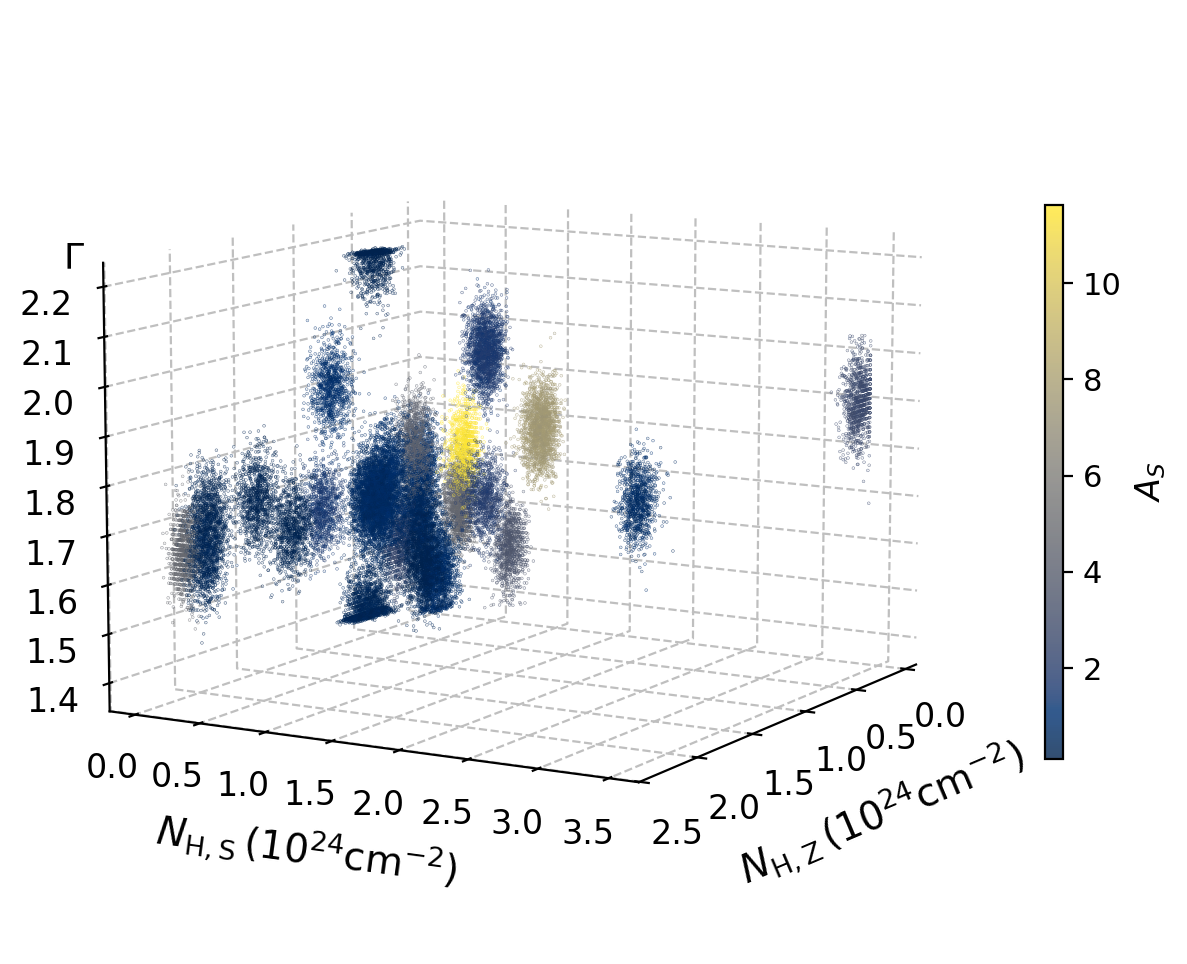}

    \caption{Literature-based \myt-decoupled values used to construct simulated spectra for NN training. Note that here, the points for AGN ESO263-13 (marked by $^{\ast}$ in \tr{tab:tab_1}), for which \as~$=0$, have been excluded. {\it Left panel}: Fitted values for physical parameters \nhz, \nhs, $\Gamma$ ($x$, $y$, $z$ axes), and \as\ (color-coded as indicated by the color bar) for 34 \nustar\ observations (24 AGN) from the literature (\tr{tab:tab_1}).  {\it Right panel}: Values for the same physical parameters obtained by assuming a Gaussian distribution around each one as explained in the text (\scr{sec-sim}), and used to construct simulated spectra for training and validation. In total, 34\,000 data points are shown.}
    \label{fig:fig_2}
\end{figure*}

\subsection{Simulation spectral set}\label{sec-sim}

As mentioned above, the limited number of only 34 fitting results from the literature represents a limitation for ML NN training. We overcome this limitation and construct a much larger observationally based simulated set of spectra for the sample group as follows.

For each original observation, we generate 1,000 variations by adding Gaussian noise (centered at zero) to the original parameter values. We empirically set the standard deviation to 10\% of each parameter's dataset mean to generate parameter variations for data augmentation. For $\Gamma$, we adopt a fixed standard deviation of 0.05. We only accept generated values that remain within the physical parameter bounds defined by the original dataset. These synthetic data points form the basis of the simulated spectral set used to train and validate the NN, significantly expanding the training sample while remaining anchored to the original observational results.

The lower panel of Figure~\ref{fig:fig_2} represents parameter values derived from the original fits, shown in the upper panel, where Gaussian noise is added to each parameter to create synthetic variations around the original observational data points. The resulting dataset therefore contains 34\,000 {\it observationally based} distinct combinations directly derived from observationally based spectral fits, although these generated sets do not include the actual 34 reported parameter sets from the literature. We refer to this grid of values, together with the associated simulated spectra (see below), as the {\it observation-based grid}, since we show later that it is significantly more consistent with the actual observed and previously fitted spectra.

We generate a grid of simulated spectra, with one artificial spectrum for each combination of physical parameters, redshift, \nhgal, and exposure time, for all 34,000 such combinations. We use {\tt fakeit} in \xspec\ {\sc v. 12.13.1}, with \nustar-specific point-source Ancillary Response and Response Matrix Files (ARF and RMF \footnote{\href{https://www.nustar.caltech.edu/page/response-files}{https://www.nustar.caltech.edu/page/response-files} 27Feb2025, version 3.}) with on-axis (offset $=$ 0) ARF corresponding to a 60-arcsec extraction region. We adopt a conservative 3–30~keV energy range (679 channels), where \nustar\ spectra are less likely to suffer from background issues. This choice is also empirically supported, as we test that it leads to robust inference of physical parameters with the ML algorithm compared to different energy ranges. It is important to note that the 34\,000 simulated spectra have a signal-to-noise ratio (S/N) representative of the observational spectra for each combination of physical parameter values, since we also sample exposure times when generating the simulation grid. To test whether our simulated spectra are indeed representative of the observational S/N, we estimate the S/N of each simulated spectrum by calculating the ratio of the average signal value to its standard deviation in a clean spectral region free from emission or absorption features \citep[see, e.g.,][]{Rosales-Ortega_2012}; in our case, we choose the range 8–15~keV.

For testing purposes, we also construct a separate grid with 34\,000 sets of parameter values sampled from a uniform distribution, with the same minimum and maximum values as for the observational sample (henceforth, {\it uniform grid}). We compare below the performance of this grid with that of the observation-based grid. Since this grid is not closely constrained by the reported fitted values but only their overall range, we do not refer to this as observationally based.

\subsection{\label{sec:samples}Sample Groups: Training, validation, and test}

We subdivide the simulated dataset into two main subsets, namely a training and a test set, containing 70\% and 30\% of the total simulated spectra and associated parameter sets, respectively. The training set is used to learn the relationships between spectra and physical parameters and to tune NN hyperparameters, such as the number of layers and parameter update rules. The test set is used to assess the performance and generalization ability of the trained NN model in reproducing the physical description of the matter around AGN. The training and test sets contain 23\,800 and 10\,200 simulated spectra, respectively.

The training dataset is divided into two subsets: 80\% and 20\%. The 80\% portion is used for the initial training phase of the model, while the remaining 20\% serves as a preliminary validation set, allowing performance to be assessed without biasing the final evaluation. The final assessment is carried out on the test set, after selecting the most effective configuration for determining the physical parameters.


\section{\label{sec-meth} Implementation, analysis, and discussion}

\subsection{\label{sec-im}Simulated spectra and univariate analysis}

We implemented a univariate analysis method to assess potential correlations between features in the simulated spectra and $z$, \nhgal, and exposure time with each physical parameter. We employ the mutual information \citep[MI,][]{6773024} criterion, which provides a non-negative measure of the nonlinear dependence between two random variables, quantifying the amount of information that can be obtained about one of them by observing the other. 

The mutual information between two discrete random variables $X$ and $Y$ is defined as:
\begin{equation}
I(X;Y) = \sum_{x \in X} \sum_{y \in Y} p(x,y) \log \left( \frac{p(x,y)}{p(x)p(y)} \right),
\end{equation}
where $p(x,y)$ is the joint probability mass function of $(X,Y)$, and $p(x)$ and $p(y)$ are the marginal distributions. This value is equal to zero if and only if the two random variables are independent; higher values (which correspond to higher entropy) indicate stronger dependence.

Since flux contributes to the fitting of physical parameters, we separately analyze how different spectral regions contribute to the determination of each \myt\ physical parameter, as well as $z$, \nhgal, and exposure time, even though these implicitly characterize the spectrum. We define six energy regions in keV: $R1 \sim [3,8)$, $R2 \sim (8-13)$, $R3 \sim (13,18)$, $R4 \sim (18, 23)$, $R5 \sim (23,28)$, and $R6 \sim (28,30]$ keV. The first region is designed to include rapid energy variations and emission lines, while the remaining regions capture intervals with similar levels of noise and intensity variation. We calculate the average counts per region for each spectrum in the validation set and compute the MI value between the mean counts in each region and the physical parameters. 

\begin{figure*}
    \centering
    \includegraphics[width=0.49\linewidth]{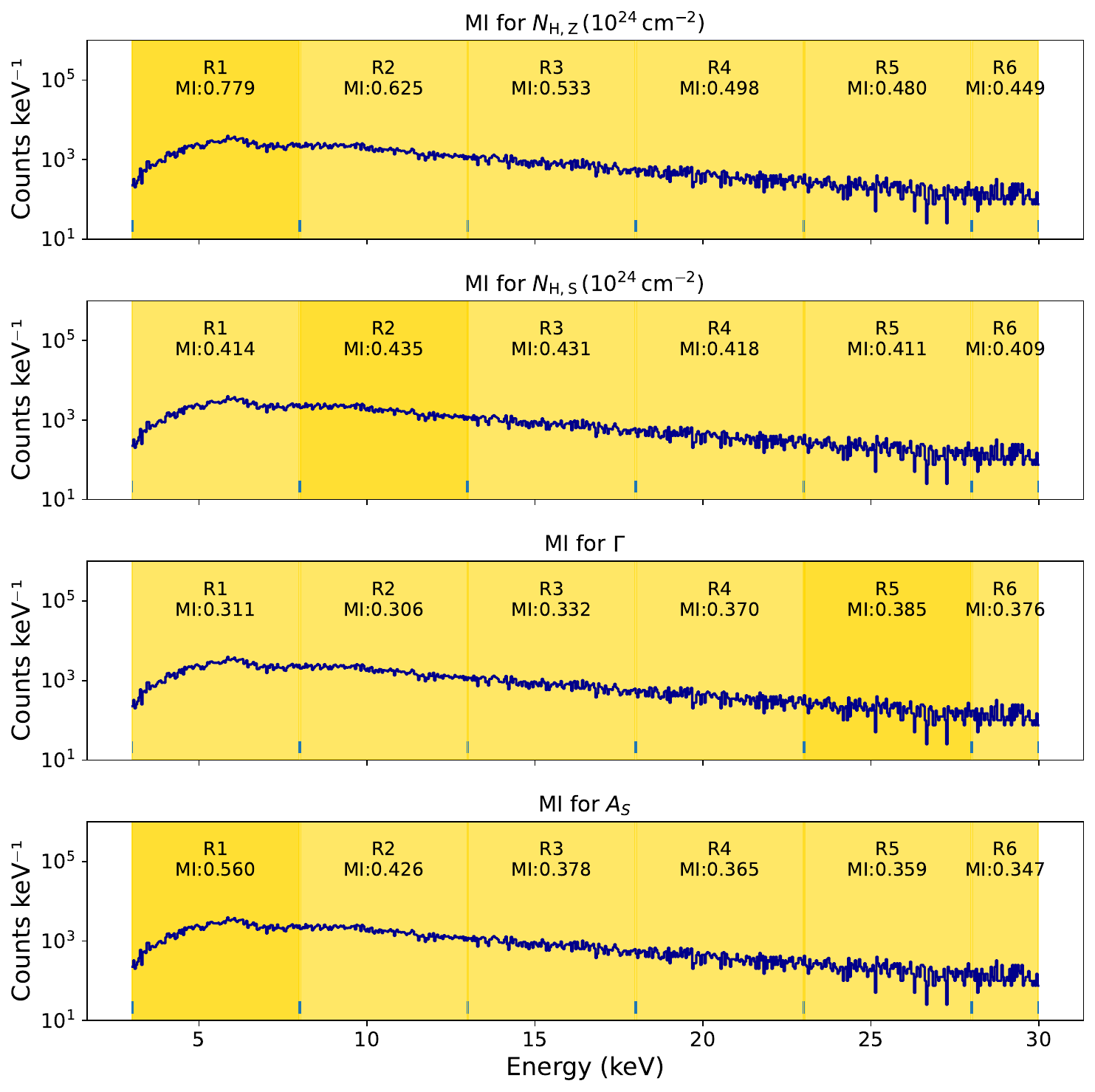}
    \includegraphics[width=0.49\linewidth]{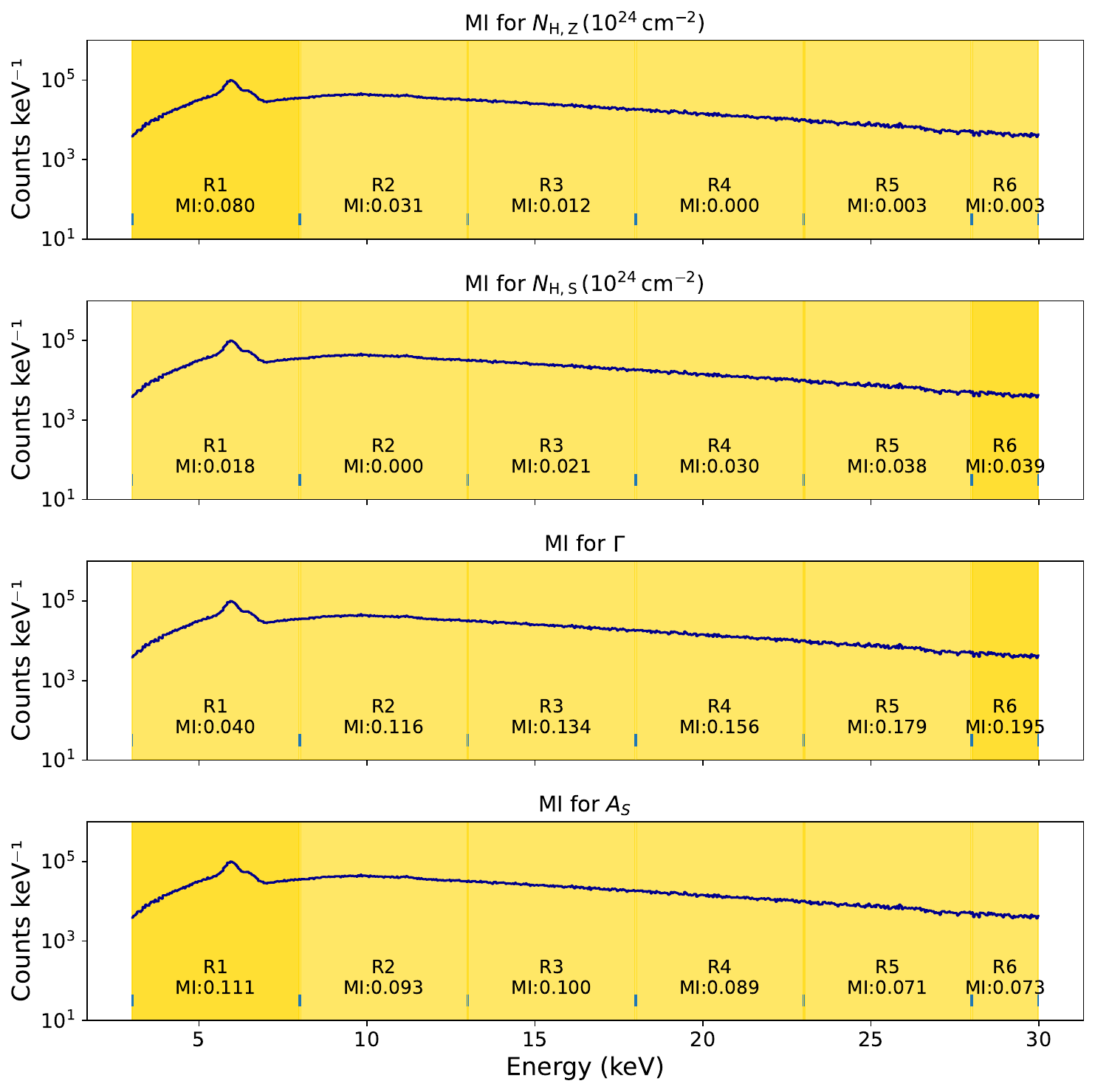}
    \caption{MI analysis between spectral regions and physical parameters. Each band shows the contribution of different energy intervals (shaded in yellow, labeled R1–R6). Each row corresponds to one of the four physical parameters: from top to bottom \nhz, \nhs, $\Gamma$, and \as. The MI value reported in each region quantifies the relative importance of that interval for determining the corresponding parameter. Left panels display spectra simulated with the observation-based grid, while right panels show the uniform grid. Note that MI = 0 indicates complete independence between the spectral region and the corresponding physical parameter.}
    \label{fig:IM}
\end{figure*}

Figure~\ref{fig:IM} shows, on the left, the spectra simulated with the observation-based grid, and on the right, those obtained with the uniform grid. In both cases, the divisions into energy regions and the MI values for each physical parameter are displayed, with the most relevant region highlighted in each case.

From this comparison, we can see that simulations from the observation-based grid achieve overall higher MI scores than those from the uniform grid, even though the latter have higher S/N. 

For the simulated spectra generated from the observation-based grid, we find that low energies (region R1) have the highest entropy for the estimation of \nhz, with an MI of 0.78. These \nhz\ MI values are consistently higher than those of the other physical parameters. They are followed by \as, with an MI above 0.56 in the same region, then by \nhs\ with an MI of 0.43 in region R2, and finally by $\Gamma$, with an MI value of 0.38. The latter becomes relevant at higher energies in region R5, with a comparable MI value also present in region R6. The higher entropy of \nhz\ in this low-energy region is not surprising: observationally, if there is sufficient \nh\ along the line of sight, a spectral turnover occurs in this region.

We find that $z$, exposure time, and \nhgal\ have no significant effect on the determination of \myt\ parameters; however, they modify the shape of the spectrum. This is further discussed in Appendix~\ref{app:spectra_univariate}.

\subsection{Training and validation}
\subsubsection{Initial training}\label{initial_training}

Two independent trainings were carried out,  one with the uniform grid and one with the observation-based grid, each incorporating the physical \myt\ parameters $\Gamma$, \nhz, \nhs, and \as, as well as $z$, \nhgal, and exposure time. All variables are scaled using the {\sc MinMaxScaler} method from the {\sc scikit-learn} library, which improves NN model performance: We obtained the best results (lowest $\mathcal{L}_{\rm{NLL}}$) using {\sc MinMaxScaler}. Each training run uses batches of 64 samples, with an internal 80\%-20\% {\sc sbi} split between training and validation, to independently monitor $\mathcal{L}_{\rm{NLL}}$ on both sets. 

Training runtimes vary depending on the grid used. On average, a complete training session with 19\,040 spectra requires approximately 4.27 minutes over 71 epochs for the observationally based training set, and 4.42 minutes over 83 epochs for the uniform grid-based training set on a {\sc CPU: [logical\_cores: 128, processor: x86\_64, RAM: 251.54 GB]}.

\begin{figure*}[!htbp]
    \centering
    \includegraphics[width=1\linewidth]{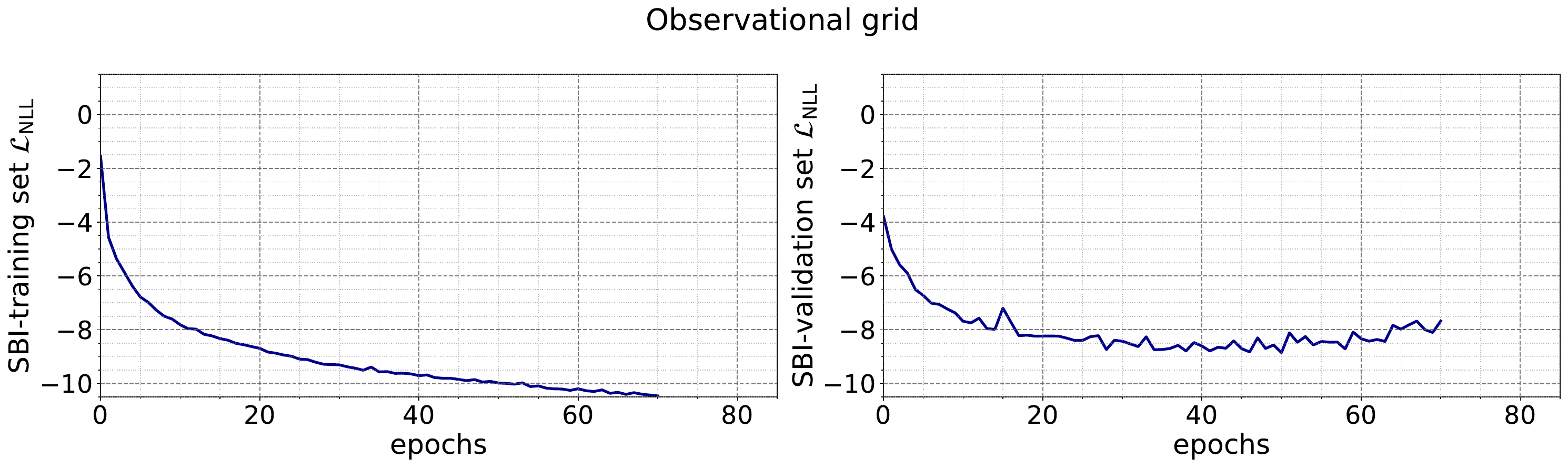}
    \includegraphics[width=1\linewidth]{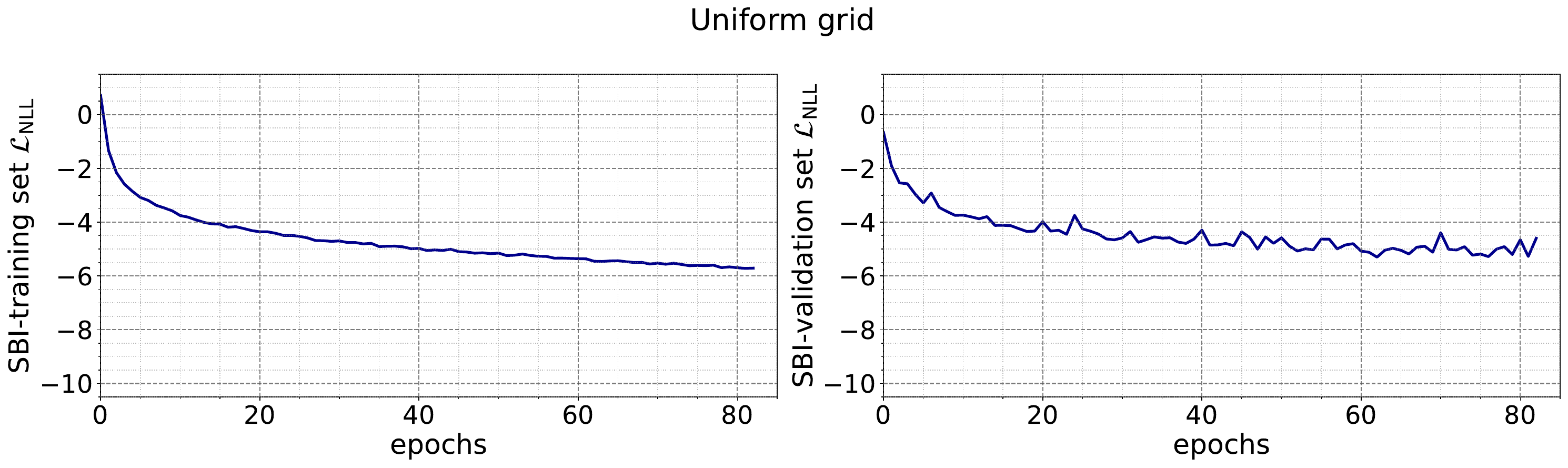}
    \caption{Training and validation loss as a function of epochs for our SBI-NPE implementation. {\it Upper panel}: simulations generated from the observation-based grid. {\it Lower panel}: simulations generated from the uniform grid.}
    \label{fig:training_curves}
\end{figure*}

In Figure~\ref{fig:training_curves}, the $\mathcal{L}_{\rm{NLL}}$ curves show that the observation-based grid achieves superior performance, with significantly lower losses (reaching -10 in training and -8 in validation) compared to the uniform grid (-5.5 and -4.5, respectively). This suggests that its sampling strategy is more effective for this specific problem. Both configurations show stable convergence without signs of overfitting, as indicated by the absence of divergence between training and validation curves, lack of excessive fluctuations, and no growing gaps that would suggest the model is memorizing the training data instead of generalizing.

\subsubsection{Validation}

We used the trained ML algorithm to make predictions for the physical parameters associated with the simulated spectra. For each spectrum in the validation set, 1\,000 posterior samples are drawn and rescaled to physical units, from which the mode (predicted value) is estimated. The true parameter values are already known, since they were used to generate the simulated spectra. Comparison between predicted and true values provides a direct assessment of the NN model's ability to recover physical parameters.

\begin{figure*}
    \centering
    \includegraphics[width=1\linewidth]{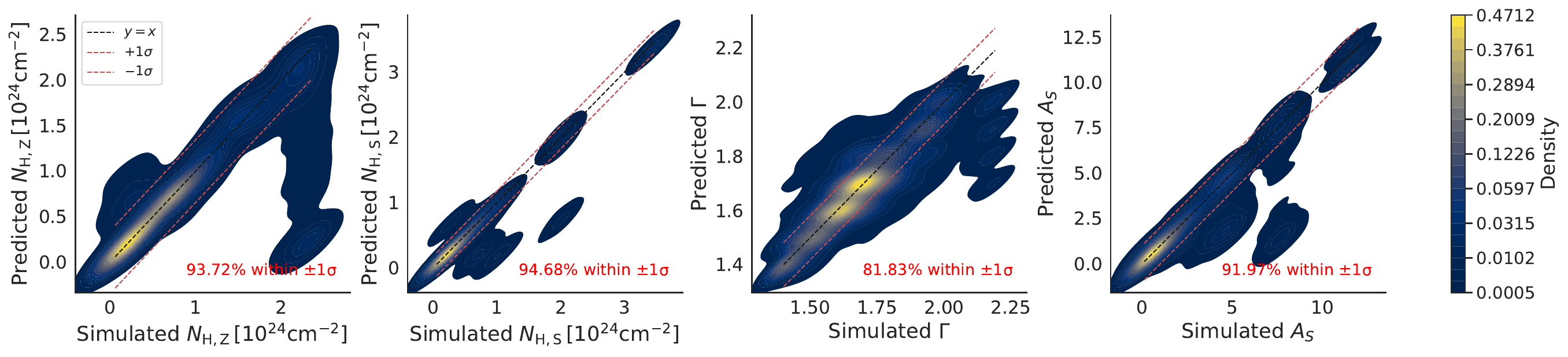}
    \includegraphics[width=1\linewidth]{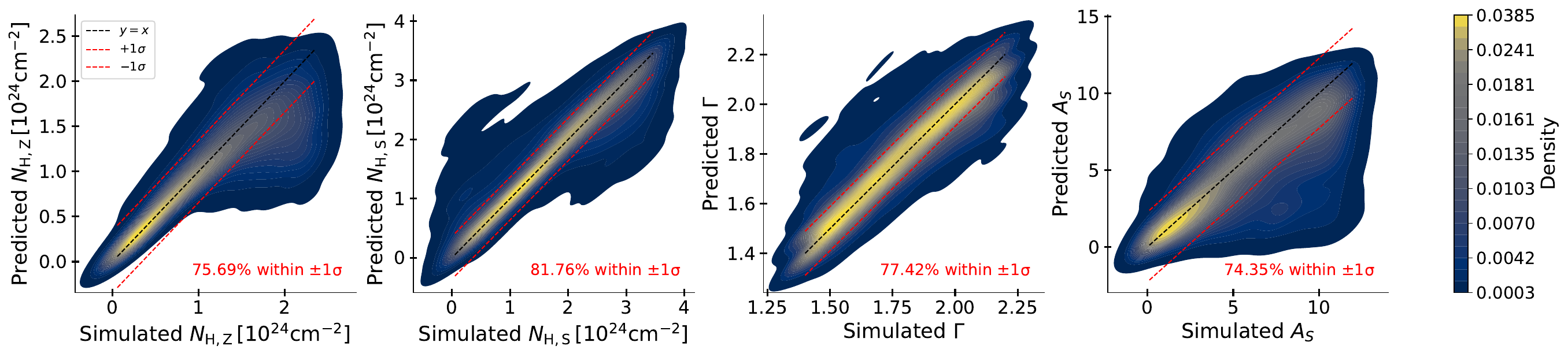}
    \caption{Predicted versus simulated values of the physical parameters in the validation set. {\it Upper panel}: results obtained with the observation-based grid of simulated spectra. {\it Lower panel}: results obtained with the uniform grid of simulated spectra. Contours indicate prediction density, with the color bar corresponding to different density levels. The black dashed line corresponds to the one-to-one relation, while red dashed lines indicate the $\pm1\sigma$ range.}
    \label{fig:pred}
\end{figure*}

Figure~\ref{fig:pred} shows the NN-predicted versus simulated (true) parameter values for both grids. The top panel corresponds to the observation-based grid, while the bottom panel corresponds to the uniform grid. Contours indicate the density of predictions, and the color represents intensity (yellow regions correspond to higher density). The black dashed line represents the one-to-one relationship, with a slope of unity suggesting a NN model with high predictive power. The red dashed lines indicate the $\pm1\sigma$ confidence region.

These results demonstrate very high predictive power for the observation-based grid, for which the NN model achieves high accuracy: 93.72\% of \nhz, 94.68\% of \nhs, 81.83\% of $\Gamma$, and 91.97\% of \as\ predictions fall within $\pm1\sigma$ of the true simulated values.  In contrast, the uniform grid exhibits a lower level of predictive power, with 75.69\% of \nhz, 81.76\% of \nhs, 77.42\% of $\Gamma$, and 74.35\% of \as\ predictions within the $\pm1\sigma$ range. While both grids show similar convergence (Figure~\ref{fig:training_curves}) during training, the consistently lower accuracy of the uniform grid during validation highlights the importance of training with realistic observational distributions.

We also analyze the joint prediction of all parameters — that is, the fraction of spectra for which all four physical parameters are simultaneously recovered within $\pm 1\sigma$, for both the observationally based and uniform grids. For the former, 74.26\% (3535/4760) of the validation sample has all parameters within $\pm 1\sigma$. For the uniform grid, the corresponding fraction drops to 40.86\% (1945/4760), indicating a substantially lower joint predictive accuracy.

As explained before, the observationally based simulated spectra faithfully represent observed \nustar\ AGN spectra across a range of physical model parameters. Therefore, these validation results indicate that the NN model and methodology have the potential to be applied—or further tested—on real observed spectra under similar conditions. This is further investigated via the case study in Section~\ref{sec-resu}. The observation-based grid generalizes well to conditions that mimic real observed spectra, despite higher noise levels.

In summary, the discussion above on training and validation shows that the choice of grid plays a decisive role in NN model performance. We therefore adopt the strategy described and train the algorithm using the observation-based grid.

\subsection{Leave-One-Seed-Out Cross-Validation}
\label{sec:loso}

The training and validation procedure provides a robust assessment of the \mlmyt\ algorithm within the parameter space covered by the full observational grid. However, it 
does not directly address the question of generalization to entirely new AGN observations whose parameter combinations were not represented in the  training set. To evaluate this more stringent condition, we implement a leave-one-seed-out (LOSO) cross-validation scheme.

The 34\,000 simulated spectra of the observation-based grid are organized into 34 groups of 1{,}000 spectra each, where every group is anchored to one of the 34 published \textit{NuSTAR} spectral fitting results (seeds) from Table~\ref{tab:tab_1}. In the LOSO scheme, we construct 34 independent training runs. In each iteration, one seed (1\,000 spectra) is withheld entirely as the test set, while the remaining 33 seeds 
(33\,000 spectra) are used for training and validation in an 80/20 split.

This procedure ensures that no spectral variation derived from the held-out observation enters the training phase in any form, directly mimicking the scenario in which the algorithm is applied to a genuinely new source.

For each of the 34 iterations, the SBI-NPE model is trained following the same procedure described in Sec.~\ref{initial_training}, including {\sc MinMaxScaler} normalization, a uniform prior defined over the scaled training parameter ranges, and a Masked Autoregressive Flow with five MADE layers. For each of the 1\,000 spectra in the held-out test set, 1\,000 posterior samples are drawn and the mode is taken as the point estimate, with uncertainty 
quantified via $\sigma_{68}$.

When \mlmyt\ is applied to a genuinely new source, the only available information is the posterior returned by the model: the true parameter value is unknown, and the user relies on $\hat{\theta}$ 
and $\sigma_{68}$ to assess the reliability of the prediction. The practical question is therefore whether $\sigma_{68}$ correctly reflects how wrong the model could be. The LOSO framework answers this in a controlled setting: since the true parameter values of the held-out seed are known, it is possible to compare directly how much the model erred against what $\sigma_{68}$ predicted. If this discrepancy is systematically large across all 34 iterations, the model underestimates its own error whenever it encounters a source absent from its training set, which is precisely the situation faced in production for any new source. In this sense, the standard 70/30 split of Sec.~\ref{sec:samples} measures how well the model interpolates within the grid, but since training and test sets are drawn from the same mixed pool, it does not simulate the real use case. The LOSO scheme does, source by source, and is therefore the relevant diagnostic for practical use of the algorithm.

To quantify this, we compute the standardized residual
\begin{equation}
    \Delta\sigma = \frac{\hat{\theta} - \theta_{\rm true}}{\sigma_{68}},
    \label{eq:deltasigma}
\end{equation}
for each prediction across all 34 held-out test sets. If $\sigma_{68}$ correctly reflects the true prediction error, these residuals should follow a standard normal distribution $\mathcal{N}(0,1)$: a mean near zero indicates the absence of systematic bias, a standard deviation near unity indicates that $\sigma_{68}$ correctly scales the true prediction errors, and approximately 5\% of samples should fall outside the $\pm2\sigma$ interval. Table~\ref{tab:loso} reports the calibration statistics for each of the four \textsc{mytorus} physical parameters.

\begin{table}[h]
\centering

\caption{Calibration statistics from the LOSO cross-validation. For each parameter we report the mean (bias; ideal $= 0$), standard deviation (uncertainty scaling; ideal $= 1$), median, and fraction of predictions outside $\pm2\sigma$ (ideal $\approx 5\%$).}
\label{tab:loso}
\begin{tabular}{lcccc}
\hline\hline
Parameter & Mean $\Delta\sigma$ & Std($\Delta\sigma$) & 
Median $\Delta\sigma$ & Outside $\pm2\sigma$ \\
\hline
$N_{\rm H,Z}$ & $-1.45$ & $4.75$ & $-0.18$ & $33.3\%$ \\
$N_{\rm H,S}$ & $-2.07$ & $9.47$ & $-0.18$ & $31.5\%$ \\
$\Gamma$       & $-0.26$ & $2.61$ & $+0.01$ & $30.3\%$ \\
$A_{\rm S}$    & $-1.13$ & $3.23$ & $-0.42$ & $32.7\%$ \\
\hline
\end{tabular}
\end{table}

The median values of $\Delta\sigma$ are close to zero for all parameters, indicating that the model predictions are nearly unbiased at the population level. The photon index $\Gamma$ shows the best overall behavior, with a 
median of $+0.01$ and the lowest standard deviation among the four parameters. However, the standard deviations of $\Delta\sigma$ are substantially larger than unity for all parameters, ranging from 2.6 for $\Gamma$ to 9.5 for $N_{\rm H,S}$, and the fractions outside $\pm2\sigma$ 
exceed 30\% for all parameters, far above the ideal 5\%. This indicates that $\sigma_{68}$ systematically underestimates the true prediction error when the model is applied to a source absent from its training set: the 
posterior intervals are too narrow relative to the actual discrepancy between predicted and true values. Since the observation-based grid produces spectra concentrated in relatively narrow regions of parameter space (Fig.~\ref{fig:fig_2}), the posterior learned from 33 seeds may not adequately represent extreme or infrequent parameter values relevant for the held-out seed. The column densities $N_{\rm H,Z}$ and $N_{\rm H,S}$ 
are the most affected, while the relative 
normalization $A_{\rm S}$ shows intermediate behavior, and $\Gamma$ is the most robustly recovered parameter under LOSO conditions.

The high per-parameter and joint accuracy obtained with the standard 70/30 split reflects the algorithm's strong 
interpolation capability within the observational grid. The LOSO analysis reveals that extrapolation to genuinely new sources introduces additional uncertainty not fully captured by $\sigma_{68}$. In practice, predictions 
are most reliable for sources whose parameter combinations lie near the center of the training distribution; for sources near the edges of the grid or in parameter regimes underrepresented in Table~\ref{tab:tab_1}, the posterior intervals should be interpreted with additional caution.

The overconfidence identified here is a property of the posterior width rather than of the model predictions, and it points to a concrete path for improvement: enlarging and diversifying the observational grid would allow the posteriors to better represent the true uncertainty for arbitrary sources. These considerations are further discussed in the context of future work in Sec.~\ref{sec-summ}.


\subsection{Final training}

After completing the initial training and validation on 23\,800 simulated spectra, we combined the initial training and validation sets to perform a final training stage and fine-tune the NN model. In other words, after the strategy definition phase, we retrained to increase the general applicability of the ML algorithm, a strategy routinely used in this field. We then applied this final ML algorithm only once to the test set to avoid any bias in the study, thus producing a robust model.

\subsection{\label{testing}Testing}

The final ML algorithm achieves $\mathcal{L}_{\rm{NLL}} = -8.80$ in 4.83 minutes. Each spectrum in the test set is sampled with 1\,000 posterior realizations using the ML algorithm to estimate the mode for each physical parameter. The uncertainty is quantified using $\sigma_{68}$. The test-set results indicate that users of this ML algorithm can expect predictive precisions of 88.13\% for \nhz, 93.61\% for \nhs, 81.66\% for $\Gamma$, and 91.84\% for \as\ within $\pm1\sigma$.

From an observer's point of view, it is of prime importance to be able to accurately predict {\it all four} parameters together. Thus, as for the validation set, we also quantify how many spectra in the test set have all four physical parameters simultaneously recovered within a given confidence interval. By definition, the $1\sigma$ region encloses approximately 68\% of the cases, $2\sigma$ about 95\%, and $3\sigma$ about 99.7\%. We find that $70.09\%$ (7149/10200) of the spectra have all parameters within $\pm 1\sigma$, $85.31\%$ (8702/10200) within $\pm 2\sigma$, and $91.90\%$ (9374/10200) within $\pm 3\sigma$. 

\begin{figure*}
    \centering
    \includegraphics[width=1\linewidth]{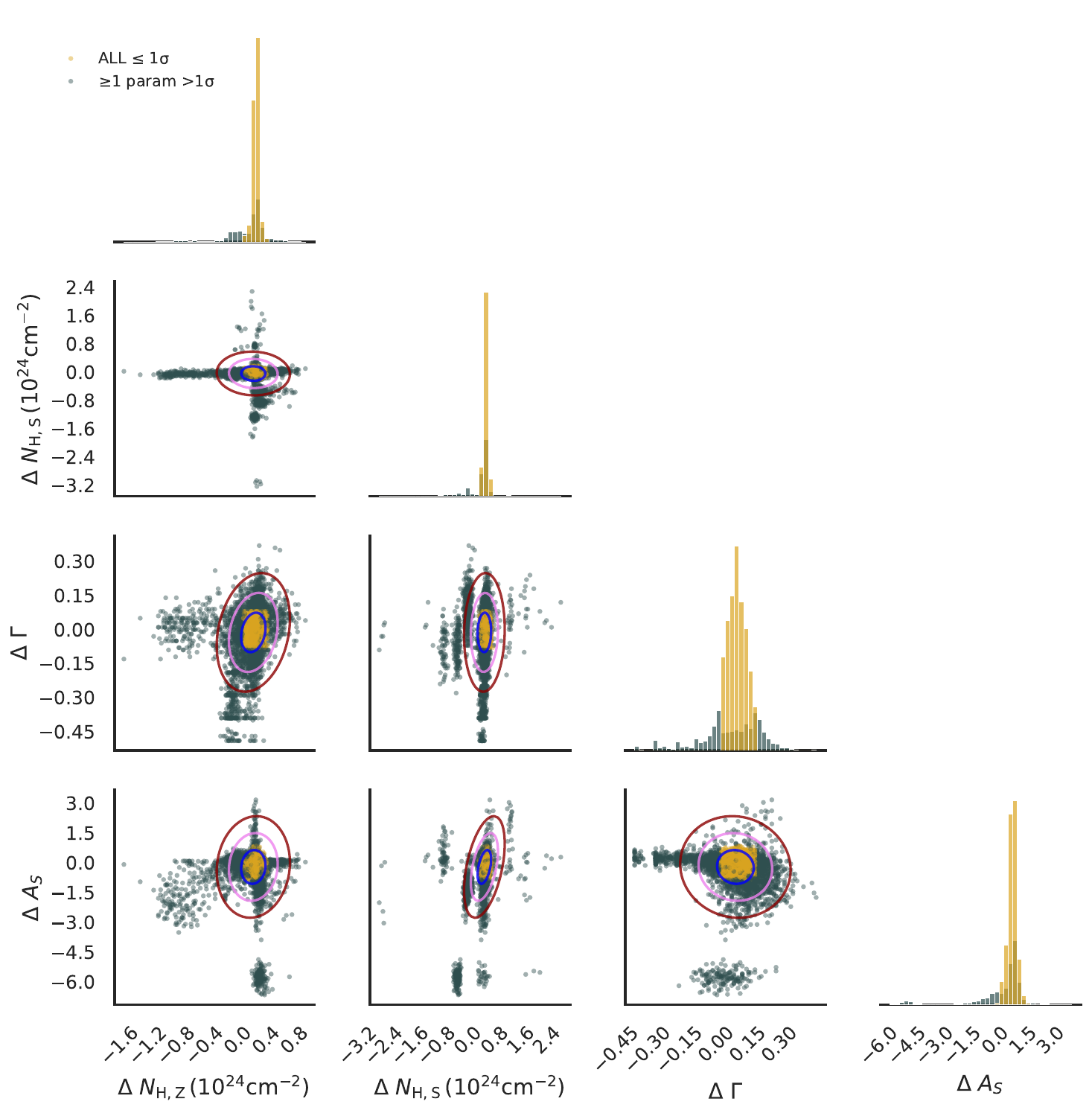}
    \caption{Pairwise distribution of residuals $\Delta = (\mathrm{prediction}\ \mathrm{value}-\mathrm{true}\ \mathrm{value})$ for the four  physical parameters: $\Delta$\nhz, $\Delta$\nhs, $\Delta\Gamma$, and $\Delta$\as\ in the test set. Off-diagonal panels show the two-dimensional residual distributions for each parameter pair, color-coded by whether all four parameters for a given spectrum fall within $1\sigma$ (yellow; ``All parameters within $1\sigma$'') or whether at least one parameter exceeds the $1\sigma$ range (teal; ``At least one parameter $>1\sigma$''). Global covariance ellipses for $1\sigma$ (blue, innermost), $2\sigma$ (violet, middle), and $3\sigma$ (burgundy, outermost) are overlaid. Diagonal panels (top left to bottom right).}
    \label{fig:pairplot_residuals}
\end{figure*}

To explore whether the errors on the physical parameters exhibit correlated structure, Fig.~\ref{fig:pairplot_residuals} presents the pairwise distributions of residuals $\Delta = (\mathrm{prediction}\ \mathrm{value}-\mathrm{true}\ \mathrm{value})$ for the four \myt\ parameters: $\Delta$\nhz, $\Delta$\nhs, $\Delta\Gamma$, and $\Delta$\as. Each off-diagonal panel shows the two-dimensional residual distribution for a parameter pair, color-coded according to whether all four parameters for that spectrum are recovered within $1\sigma$  or whether at least one parameter lies outside the $1\sigma$ range. Global covariance ellipses corresponding to $1\sigma$ (blue), $2\sigma$ (violet), and $3\sigma$ (burgundy) are overlaid to highlight the joint residual structure and the directions of maximum dispersion, while the top-left to bottom-right diagonal panels display the one-dimensional residual histograms.

Figure~\ref{fig:pairplot_residuals} indicates that the residuals are centered close to zero for all parameters, with the highest-density regions tightly clustered around the origin, supporting the absence of systematic biases and indicating that the ML algorithm generally succeeds producing small errors. Both the contours and the distribution of points suggest that the variables are weakly correlated with each other, as the ellipses are generally round (without a pronounced slope), although there may be a slight correlation or asymmetric dispersion in some pairs:  For $\Delta\Gamma$ vs.~$\Delta$\nhz, the red ellipse appears slightly more elongated and with a positive slope.  For $\Delta$\as\ vs.~$\Delta\Gamma$, the distribution of points has a somewhat slanted or diffuse shape, although the central contours are tight.

We make this final ML algorithm, \mlmyt, publicly available at~\href{https://github.com/vanedaza/viva-streamlit}{https://github.com/vanedaza/viva-streamlit}~\cite{mlmytorus_github}. We also make available an easy-to use web-based version, \webmlmyt, which can be accessed at \href{https://web-ml-mytorus.streamlit.app/}{https://web-ml-mytorus.streamlit.app/}~\cite{mlmytorus_webapp}. Here, one can upload a two-column text file, where the first column represents energy in keV and the second counts keV$^{-1}$ for \nustar\ spectra, to obtain in seconds \mlmyt-based posterior values with associated uncertainties for the four \myt-decoupled physical parameters: \nhz, \nhs, $\Gamma$, and \as.

\section{Case study}\label{sec-resu}

The algorithm and its publicly available interface were developed to enable users to infer the \myt\ model parameters in \emph{decoupled mode} for any observed \nustar\ AGN spectrum. To demonstrate \mlmyt's functionality and compare it with a manual fit performed in \xspec, we apply it to a single \nustar\ observation of the well-known Seyfert~2 galaxy NGC~4388 (ObsID:~60061228002), with an exposure time of 21.4~ks, a Galactic column density of \nhgal~$=2.57\times10^{20}$~cm$^{-2}$, and redshift $z=0.0086$. This observation was previously analyzed by \cite{Torres-Alba_2023} but with a different configuration of the \myt\ model. Here, we reanalyze the data to enable a direct comparison between a traditional \xspec\ fit and our ML-based \mlmyt\ approach.

The \mlmyt\ framework requires as input a calibrated, background-subtracted source spectrum. We reduced \nustar\ FPMA and FPMB data using the \nustar\ Data Analysis Software {\sc NuSTARDAS v2.1.5}, and the \nustar\ calibration database v20250812. We ran \texttt{nupipeline} to produce screened and calibrated level-2 event lists, and \texttt{nuproducts} to extract source and background spectra, and corresponding responses. A circular, 130 arcsec-radius region centered on the source was used to generate the source data. Another circular region, offset from the source and with a radius of 180 arcsec, was employed to extract the background. The ARF and RMF files were generated with the {\sc numkarf} and {\sc numkrmf} tasks, respectively. The resulting observation has an S/N ratio of 1.01, which lies at the lower end of the range covered by the training and validation distributions of our neural posterior estimator (see Fig.~\ref{fig:hist_sn}).

We used \webmlmyt\ for the combined FPMA $+$ FPMB spectrum (without any energy cuts), uploaded as a two-column plain text file containing energy (keV) and counts keV$^{-1}$ (obtained with command {\tt plot counts} in \xspec). The input parameters were \nhgal, $z$, and a total exposure time of $\sim$42.8~ks (i.e., twice the exposure time for a single detector). The application outputs the posterior distributions for the four physical parameters \nhz, \nhs, $\Gamma$, and \as, from which we extract the posterior \emph{mode} and the 68\% and 90\% credible intervals (central p16–p84 and p5–p95, respectively). For comparison purposes, we performed a manual \xspec\ fit, using the \myt-decoupled model setup presented in \scr{sec-mytorus}. The fitting was done ``blindly," without consulting the output of \webmlmyt.

\begin{figure}
    \centering
    \includegraphics[width=1\linewidth]{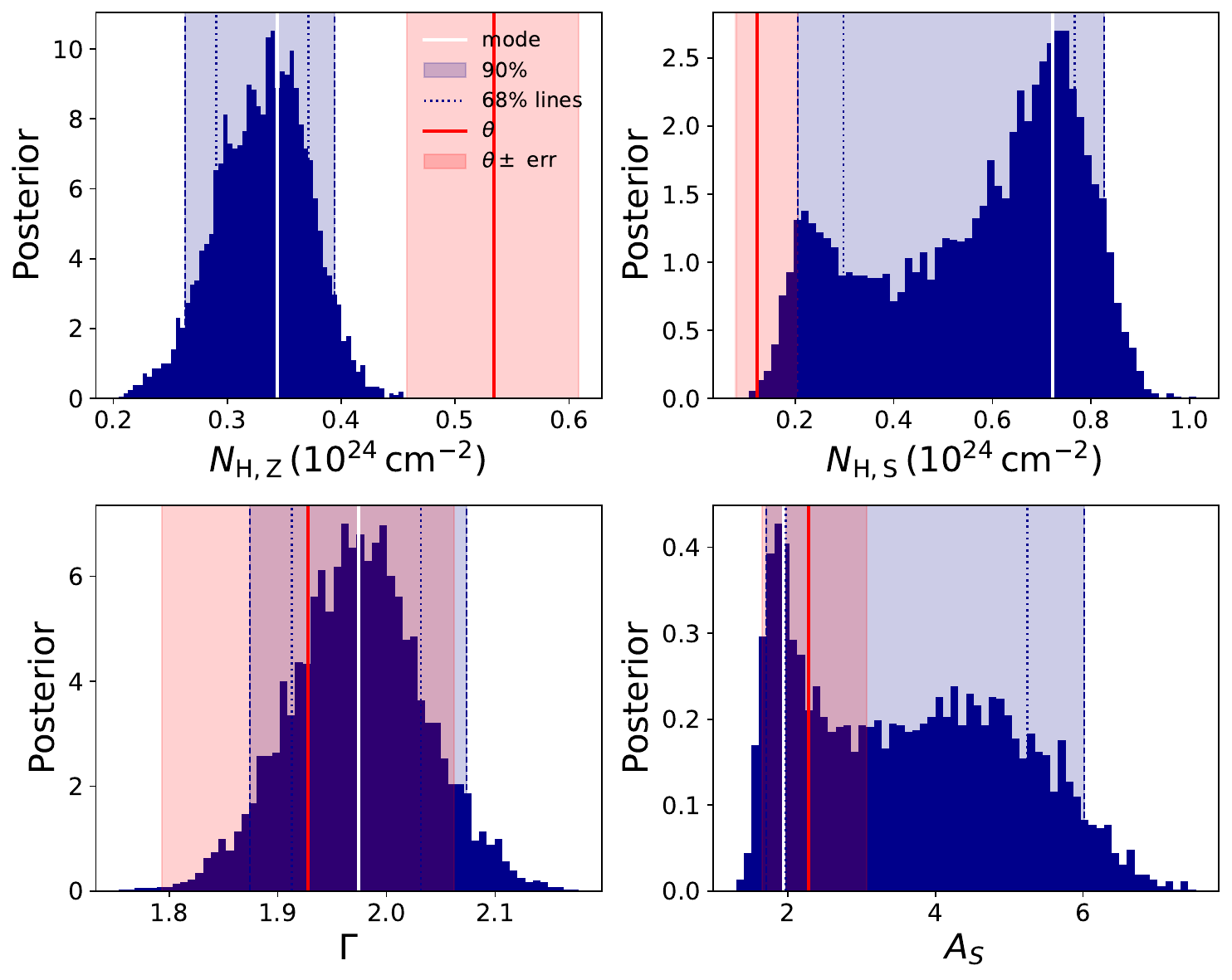}
    \caption{Posterior histograms for \nhz, \nhs, $\Gamma$, and \as\ from \mlmyt. The white vertical line marks the posterior mode predicted by the neural network. The dark–blue shaded band shows the 90\% credible interval (p5–p95), and the dotted dark–blue lines mark the 68\% credible limits (p16 and p84). The red vertical line indicates the \xspec\ best-fit value from \myt\ (decoupled mode), and the semi-transparent red band shows its 90\% confidence interval.}
    \label{fig:3d}
\end{figure}

Table~\ref{tab:comparison} (see also \fr{fig:3d}) summarizes the comparison between parameter posterior estimates obtained with \mlmyt\ and those obtained for \myt\ decoupled via a standard \xspec\ fit. The 90\% confidence intervals correspond to $\Delta C=2.706$ for the \xspec\ fit and to the p5-p95 posterior quantiles for \mlmyt.

\begin{table*}
\centering
\caption{Comparison between \mlmyt\ posterior modes and \xspec\ \myt-decoupled best-fit values for NGC~4388. The 90\% intervals correspond to p5–p95 for \mlmyt\ and $\Delta C=2.706$ for \xspec.}
\label{tab:comparison}
\begin{tabular}{lcccc}
\hline
Parameter & 
\mlmyt\ mode & 
\mlmyt\ 90\% & 
\xspec\ best-fit & 
\xspec\ 90\% (ΔC)\\ 
\hline
\nhz\ ($10^{24}$ \cunits) & 0.336 & [0.260, 0.396] & 0.534 & [0.458, 0.609] \\
\nhs\ ($10^{24}$ \cunits) & 0.725 & [0.207, 0.833] & 0.123 & [0.080, 0.205] \\
$\Gamma$           & 1.964 & [1.873, 2.074] & 1.928  & [1.794, 2.062] \\
\as             & 1.937 & [1.707, 6.080] & 2.283 & [1.650, 3.066] \\
\hline
\end{tabular}
\end{table*}

The results of this comparison provide very useful insight into the intricacies of constraining these model parameters. First, within the 90\% uncertainty range both methods obtain consistent values for the photon index and the scattered component normalization, with $\Gamma^{\rm ML}=1.96^{+0.11}_{-0.09}$, $\Gamma^{\xspec}=1.93^{+0.14}_{-0.14}$, \as$^{\rm ML}=1.94^{+4.14}_{-0.23}$, and \as$^{\xspec}=2.28^{+0.78}_{-0.64}$. In contrast, the largest discrepancies arise in the column densities. For the line-of-sight component, \nhz$^{\xspec}=0.53^{+0.08}_{-0.07}\times10^{24}\,\mathrm{cm^{-2}}$ is higher than the \mlmyt\ value of \nhz$^{\rm ML}=0.34^{+0.06}_{-0.08}\times10^{24}\,\mathrm{cm^{-2}}$, while for the global column density the opposite trend is seen: \nhs$^{\rm ML}=0.72^{+0.11}_{-0.52}\times10^{24}\,\mathrm{cm^{-2}}$ is larger than \nhs$^{\xspec}=0.12^{+0.08}_{-0.04}\times10^{24}\,\mathrm{cm^{-2}}$.

While this is a formal discrepancy, in fact it provides insight into an issue that is arguably unavoidable in at least some cases with data of \nustar-level spectral resolution. Looking at the upper right panel of Fig.~8, the ML-reported most probable \nhs\ solution essentially represents the hightest peak in the posterior distribution. There is, in fact, a second peak at \sss\ten{0.2}{24}\cunits\ which, within {\it its} 90\%\ uncertainty overlaps with the \xspec\ formally reported best-fit \nhs\ solution. Similarly, further exploration of parameter space within \xspec\ shows that there is a second solution for which {\it all} parameter values agree within 90\%\ with the reported ML solution. This second solution corresponds to $C=695.76$ (c.f. 690.45 for the formal best fit) for 674 bins. This is a change in $C$ of $<$1\%, which, given uncertainties due to resolution, noise, or exposure time, can hardly be considered significant. In other words, the data are consistent with two, degenerate solutions, and this is picked up both by the ML algorithm and by standard frequentist \xspec\ fitting.

Note further that while the validation and testing stages showed that in the limit of large samples both individual parameters, as well as all four parameters simultaneously, are recovered via the ML algorithm with high predictive accuracy, the case study discussed here deals with a {\it single} spectrum chosen at random. It is an inevitable statistical fact that there can be no guarrantee that such a choice will produce a case that will always correspond to the best prediction for all four parameters. However, the power of the validation and testing results is that they provide one with a quantitative, robust estimate of the probability that such a single-spectrum based prediction is correct, a probability that, as we have already shown, is significantly high for the observationally-based version of the ML algorithm that we make publicly available. In addition, the case study is a good example for the general consideration of what the ML approach does for degenerate or nearly degenerate solutions, how the data analyst can be alerted that this may be happening, and what the course of action should be. We have shown that the ML algorithm produces quantitative posterior distributions that can alert the user of the existence of degenerate solutions. Such cases should be noted and reported accordingly. Thus, overall, rather than suggesting a weakness, we consider the case study to actually indicate the robustness and usefulness of the ML algorithm for individual spectra.


\section{Summary and Conclusions}\label{sec-summ}

In this work, we developed a Simulation-Based Inference (SBI) framework using Neural Posterior Estimation (NPE) to infer the four key physical parameters of AGN \x\ spectra within the \myt\ decoupled model setup. Trained and tested on 34\,000 simulated spectra derived from observationally based parameter distributions of 24 AGN, the resulting \mlmyt\ algorithm achieves high predictive accuracy and robustness as evidenced via validation and testing. We also demonstrate the application of the method to real data using a \nustar\ observation of NGC~4388. Formally, only the predictions for $\Gamma$ and \as\ closely match the \xspec\ best-fit results, whereas \nhz\ and \nhs\ differ. However, this reflects the intrinsic bimodality of the \myt\ model for a certain regime of the equivalent width of the narrow \feka\ line, which allows two statistically almost equivalent solutions for the column densities (see, for example, Figure 8 in \cite{murphy2009}). While the \mlmyt\ algorithm converges toward one of these minima, \xspec\ may settle on the other depending on the shape of the likelihood landscape. Thus, for cases in which bimodality is present, \mlmyt\ provides a posterior histogram of \nhs\  that may indicate two possible values, which can then be used to guide the search for another local minimum with \xspec.

\begin{enumerate}
    \item {\it Sample:} We compile a set of 34 published \nustar\ observations previously analyzed with the \myt\ model in decoupled mode. We thus obtain a unified view of how the four main physical parameters are related in a multidimensional space (Table~\ref{tab:tab_1}, Fig.~\ref{fig:fig_2}). This sample is representative of AGN modeled with decoupled \myt\ in the literature, although it is not intended to be statistically complete.

    \item {\it Feature selection:} Using the mutual information (MI) criterion, we identified the most informative spectral regions for constraining each physical parameter of the \myt\ model and analyzed how exposure time, \nhgal, and redshift ($z$) correlate in relation to these features (Sect.~\ref{sec-im}).

    \item {\it Observational vs uniform grid:} We compared the training performance obtained using a parameter grid based on observationally motivated distributions with that from a uniformly sampled grid, showing that the observation-based grid provides higher accuracy and greater stability (Sect.~\ref{initial_training}).

    \item {\it Precision:} Agreement between predicted and simulated values for \nhz, \nhs, $\Gamma$, and \as\ is high, with deviations comparable to typical uncertainties in \xspec\ fits. Beyond the per-parameter accuracy, the joint recovery of all four physical parameters is also strong: in the test set, 70.09\% of the spectra have all parameters within $\pm 1\sigma$, 85.31\% within $\pm 2\sigma$, and 91.90\% within $\pm 3\sigma$. These results demonstrate that the network not only performs well individually for each parameter but also maintains consistency across the full multidimensional parameter space.

    \item {\it Computational cost:} The algorithm was trained on a CPU in just a few minutes, using a moderate number of simulations generated within a few hours, highlighting both its speed and low computational cost.
    
    \item {\it Reproducibility:} The \mlmyt\ algorithm is publicly available on GitHub at \href{https://github.com/vanedaza/viva-streamlit}{https://github.com/vanedaza/viva-streamlit}~\cite{mlmytorus_github}, ensuring full result reproducibility and transparency (Sect.~\ref{testing}).

    \item {\it Web platform:} For \mlmyt\ we further make publicly available a web interface, \webmlmyt, at \href{https://web-ml-mytorus.streamlit.app}{https://web-ml-mytorus.streamlit.app}~\cite{mlmytorus_webapp}, that allows users to upload a \nustar-quality energy spectrum, input \nhgal, $z$, and exposure time, and obtain the posterior \myt\ physical parameters within seconds and with accuracy comparable to traditional \myt\ fits (Sect.~\ref{testing}).

    \item {\it Application to observations:} Detailed testing shows that the method can recover physical parameters consistent with traditional fits for sources whose spectra fall within the training domain. In a random case study, column-density posterior values appear formally discrepant from values obtained with \xspec, however we find that \xspec\ affords a second, statistically close solution, that is similar to the one obtained with \mlmyt, due to the intrinsic data bimodality.

    \item {\it Scope and limitations:} 
    \mlmyt\ is trained on spectra whose corresponding physical parameter values lie within a parameter-space region spanned by a 34000-strong grid, ultimately originating in a set of 34 fitted observed spectra. Since the training labels essentially come from \xspec\ fits, systematic uncertainties and biases associated with the solutions in these fits will naturally propagate to the ML predictions. The algorithm’s validity limits are thus clear: Predictions should be most reliable for spectra with physical parameters within or near the training ranges ($N_{\mathrm{H,Z}}\!\in$ [0.07–2.34]$\times10^{24}$~cm$^{-2}$, $N_{\mathrm{H,S}}\!\in$ [0.14–6.49]$\times10^{24}$~cm$^{-2}$, $\Gamma\!\in$ [1.47–2.26], $A_S\!\in$ [0.10–4.77], \nhgal$\,\in[7.5\times10^{19}$–$2.6\times10^{21}]$~cm$^{-2}$, $z\!\in$ [0.004–0.107], S/N $\approx$ 1–6). Performance may degrade for extreme or out-of-range cases (e.g., heavily Compton-thick, unobscured type 1, high-z, or very low or high S/N).

    {\item  {\it Generalization caveat}: In spite of the successful case study demonstration, we caution that a detailed leave-one-seed-out test indicates that the ML algorithm may only generalize well for observed spectra that are representative of the parameter ranges and combinations used to train the algorithm. Generalization beyond such cases is not demonstrated.}

   \item {\it Future work:} Extending \mlmyt\ to new regimes will require expanding the training grid or incorporating reconditioning approaches such as Simformer \citep{Gloeckler_2024}. Including moderate variations of spectral-broadening parameters or other high-entropy spectral features in the simulations may help the model better capture mixed or degenerate regimes and further improve robustness. Further extensions can include ML training for different \x\ telescopes and instruments, combinations of instruments, as well as distant reflection spectra from other \x–emitting accreting compact sources such as Galactic \x\ binaries, which have also been modeled with spectral models such as \myt. ML applications to other physical spectral models is a further obvious extension.

\end{enumerate}

Taken together, these results show that, for the four physical parameters used here and \nustar-quality spectra, \mlmyt\ provides a fast, accurate, and fully reproducible alternative to conventional \myt\ fitting in \xspec, with the added benefit of a public web interface that returns physically consistent estimates within seconds. Its applicability to new sources can be strengthened by expanding parameter-space coverage and spectral-model diversity in training. 

\begin{acknowledgements}
We thank an anonymous referee for insightful comments that helped improve the quality of the manuscript. We thank the group of J{\"o}rn Wilms at Dr. Karl Remeis-Observatory (Astronomical Institute of the University of Erlangen-Nuremberg) for their support with data hosting and reduction, as well as useful discussions. 
This research has made use of the \nustar\ Data Analysis Software ({\sc NuSTARDAS}), jointly developed by the ASI Space Science Data Center (SSDC, Italy) and the California Institute of Technology (Caltech, USA). I.V.D.P., P.T., and V.M.F. acknowledge support from NASA grant No. 80NSSC22K0411 (solicitation NNH21ZDA001N-ADAP, PI P.~Tzanavaris). This work is supported by NASA under the CRESST cooperative agreement award number 80GSFC24M0006. This research made use of Python and open-source scientific software, including {\sc NumPy}, {\sc pandas}, {\sc Matplotlib}, {\sc PyTorch}, and {\sc sbi}. X-ray spectral simulations and model evaluations were performed using {\sc XSPEC}. The accompanying web application was developed using {\sc Streamlit}.
\end{acknowledgements}
\clearpage
\bibliographystyle{apsrev4-2}
\bibliography{apssamp}

@ARTICLE{ghisellini1994,
       author = {{Ghisellini}, G. and {Haardt}, F. and {Matt}, G.},
        title = "{The contribution of the obscuring torus to the X-ray spectrum of Seyfert galaxies: a test for the unification model.}",
      journal = {\mnras},
         year = 1994,
        month = apr,
       volume = {267},
        pages = {743-754},
          doi = {10.1093/mnras/267.3.743},
archivePrefix = {arXiv},
       eprint = {astro-ph/9401044},
 primaryClass = {astro-ph},
       adsurl = {https://ui.adsabs.harvard.edu/abs/1994MNRAS.267..743G}
}

@ARTICLE{Matt1996,
       author = {{Matt}, G. and {Fabian}, A.~C. and {Ross}, R.~R.},
        title = "{Iron K fluorescent lines from relativistic, ionized discs}",
      journal = {\mnras},
         year = 1996,
        month = feb,
       volume = {278},
       number = {4},
        pages = {1111-1120},
          doi = {10.1093/mnras/278.4.1111},
       adsurl = {https://ui.adsabs.harvard.edu/abs/1996MNRAS.278.1111M}
}

@ARTICLE{sulentic1998,
   author = {{Sulentic}, J.~W. and {Marziani}, P. and {Zwitter}, T. and {Calvani}, M. and 
	{Dultzin-Hacyan}, D.},
    title = "{On the Origin of Broad Fe K{$\alpha$} and H I H{$\alpha$} Lines in Active Galactic Nuclei}",
  journal = {\apj},
   eprint = {astro-ph/9712336},
     year = 1998,
    month = jul,
   volume = 501,
    pages = {54-68},
      doi = {10.1086/305795},
   adsurl = {http://adsabs.harvard.edu/abs/1998ApJ...501...54S}
}

@INPROCEEDINGS{Gelbord2001,
       author = {{Gelbord}, J. and {Weaver}, K.~A.},
        title = "{Variable Fe-K{\ensuremath{\alpha}} line profiles in Sy 1 galaxies}",
    booktitle = {X-ray Astronomy: Stellar Endpoints, AGN, and the Diffuse X-ray Background},
         year = 2001,
       editor = {{White}, Nicholas E. and {Malaguti}, Giuseppe and {Palumbo}, Giorgio G.~C.},
       series = {American Institute of Physics Conference Series},
       volume = {599},
        month = dec,
    publisher = {AIP},
        pages = {622-625},
          doi = {10.1063/1.1434701},
       adsurl = {https://ui.adsabs.harvard.edu/abs/2001AIPC..599..622G}
}

@ARTICLE{weaver2001,
   author = {{Weaver}, K.~A. and {Gelbord}, J. and {Yaqoob}, T.},
    title = "{Variable Iron K{$\alpha$} Lines in Seyfert 1 Galaxies}",
  journal = {\apj},
   eprint = {astro-ph/0008522},
     year = 2001,
    month = mar,
   volume = 550,
    pages = {261-279},
      doi = {10.1086/319713},
   adsurl = {http://adsabs.harvard.edu/abs/2001ApJ...550..261W}
}

@ARTICLE{yaqoob2001,
   author = {{Yaqoob}, T. and {George}, I.~M. and {Nandra}, K. and {Turner}, T.~J. and 
	{Serlemitsos}, P.~J. and {Mushotzky}, R.~F.},
    title = "{Physical Diagnostics from a Narrow Fe K{$\alpha$} Emission Line Detected by Chandra in the Seyfert 1 Galaxy NGC 5548}",
  journal = {\apj},
   eprint = {astro-ph/0008471},
     year = 2001,
    month = jan,
   volume = 546,
    pages = {759-768},
      doi = {10.1086/318315},
   adsurl = {http://adsabs.harvard.edu/abs/2001ApJ...546..759Y}
}

@ARTICLE{yaqoob2004,
   author = {{Yaqoob}, T. and {Padmanabhan}, U.},
    title = "{The Cores of the Fe K Lines in Seyfert 1 Galaxies Observed by the Chandra High Energy Grating}",
  journal = {\apj},
   eprint = {astro-ph/0311551},
     year = 2004,
    month = mar,
   volume = 604,
    pages = {63-73},
      doi = {10.1086/381731},
   adsurl = {http://adsabs.harvard.edu/abs/2004ApJ...604...63Y}
}

@ARTICLE{nandra2006,
   author = {{Nandra}, K.},
    title = "{On the origin of the iron K{$\alpha$} line cores in active galactic nuclei}",
  journal = {\mnras},
   eprint = {astro-ph/0602081},
     year = 2006,
    month = may,
   volume = 368,
    pages = {L62-L66},
      doi = {10.1111/j.1745-3933.2006.00158.x},
   adsurl = {http://adsabs.harvard.edu/abs/2006MNRAS.368L..62N}
}

@ARTICLE{shu2010,
   author = {{Shu}, X.~W. and {Yaqoob}, T. and {Wang}, J.~X.},
    title = "{The Cores of the Fe K{$\alpha$} Lines in Active Galactic Nuclei: An Extended Chandra High Energy Grating Sample}",
  journal = {\apjs},
archivePrefix = "arXiv",
   eprint = {1003.1790},
 primaryClass = "astro-ph.HE",
     year = 2010,
    month = apr,
   volume = 187,
    pages = {581-606},
      doi = {10.1088/0067-0049/187/2/581},
   adsurl = {http://adsabs.harvard.edu/abs/2010ApJS..187..581S}
}

@ARTICLE{shu2011,
   author = {{Shu}, X.~W. and {Yaqoob}, T. and {Wang}, J.~X.},
    title = "{Chandra High-energy Grating Observations of the Fe K{$\alpha$} Line Core in Type II Seyfert Galaxies: A Comparison with Type I Nuclei}",
  journal = {\apj},
archivePrefix = "arXiv",
   eprint = {1107.0195},
     year = 2011,
    month = sep,
   volume = 738,
      eid = {147},
    pages = {147},
      doi = {10.1088/0004-637X/738/2/147},
   adsurl = {http://adsabs.harvard.edu/abs/2011ApJ...738..147S}
}

@ARTICLE{zoghbi2019,
       author = {{Zoghbi}, A. and {Miller}, J.~M. and {Cackett}, E.},
        title = "{Revisiting the Spectral and Timing Properties of NGC 4151}",
      journal = {\apj},
         year = 2019,
        month = oct,
       volume = {884},
       number = {1},
          eid = {26},
        pages = {26},
          doi = {10.3847/1538-4357/ab3e31},
archivePrefix = {arXiv},
       eprint = {1908.09862},
 primaryClass = {astro-ph.HE},
       adsurl = {https://ui.adsabs.harvard.edu/abs/2019ApJ...884...26Z}
}

@ARTICLE{marinucci2013,
       author = {{Marinucci}, A. and {Miniutti}, G. and {Bianchi}, S. and {Matt}, G. and {Risaliti}, G.},
        title = "{A Chandra view of the clumpy reflector at the heart of the Circinus galaxy}",
      journal = {\mnras},
         year = 2013,
        month = dec,
       volume = {436},
       number = {3},
        pages = {2500-2504},
          doi = {10.1093/mnras/stt1759},
archivePrefix = {arXiv},
       eprint = {1309.4456},
 primaryClass = {astro-ph.CO},
       adsurl = {https://ui.adsabs.harvard.edu/abs/2013MNRAS.436.2500M}
}

@ARTICLE{guainazzi2016,
  author = {{Guainazzi}, M. and {Risaliti}, G. and {Awaki}, H. and {Arevalo}, P. and
            {Bauer}, F.~E. and others},
  title = "{The nature of the torus in the heavily obscured AGN Markarian 3: an X-ray study}",
  journal = {\mnras},
  archivePrefix = "arXiv",
  eprint = {1605.02467},
  primaryClass = "astro-ph.HE",
  year = 2016,
  month = aug,
  volume = 460,
  pages = {1954--1969},
  doi = {10.1093/mnras/stw1033},
  url = {https://ui.adsabs.harvard.edu/abs/2016MNRAS.460.1954G/abstract}
}

@ARTICLE{marinucci2017,
  author = {{Marinucci}, A. and {Bianchi}, S. and {Fabbiano}, G. and {Matt}, G. and
            {Risaliti}, G. and others},
  title = "{Spatially resolved Fe K spectroscopy of NGC 4945}",
  journal = {\mnras},
  year = 2017,
  month = oct,
  volume = {470},
  number = {4},
  pages = {4039--4047},
  doi = {10.1093/mnras/stx1551},
  archivePrefix = {arXiv},
  eprint = {1706.06362},
  primaryClass = {astro-ph.HE},
  url = {https://ui.adsabs.harvard.edu/abs/2017MNRAS.470.4039M/abstract}
}

@ARTICLE{fabbiano2017,
  author = {{Fabbiano}, G. and {Elvis}, M. and {Paggi}, A. and {Karovska}, M. and
            {Maksym}, W.~P. and others},
  title = "{Discovery of a Kiloparsec Extended Hard X-Ray Continuum and Fe-K{\ensuremath{\alpha}} from the Compton Thick AGN ESO 428-G014}",
  journal = {\apjl},
  year = 2017,
  month = jun,
  volume = {842},
  number = {1},
  eid = {L4},
  pages = {L4},
  doi = {10.3847/2041-8213/aa7551},
  archivePrefix = {arXiv},
  eprint = {1705.10680},
  primaryClass = {astro-ph.HE},
  url = {https://ui.adsabs.harvard.edu/abs/2017ApJ...842L...4F/abstract}
}

@ARTICLE{xrism2024,
  author = {{XRISM Collaboration} and others},
  title = "{XRISM Spectroscopy of the Fe K{\ensuremath{\alpha}} Emission Line in the Seyfert Active Galactic Nucleus NGC 4151 Reveals the Disk, Broad-line Region, and Torus}",
  journal = {\apjl},
  year = 2024,
  month = sep,
  volume = {973},
  number = {1},
  eid = {L25},
  pages = {L25},
  doi = {10.3847/2041-8213/ad7397},
  archivePrefix = {arXiv},
  eprint = {2408.14300},
  primaryClass = {astro-ph.HE},
  url = {https://ui.adsabs.harvard.edu/abs/2024ApJ...973L..25X/abstract}
}

@ARTICLE{barret2023,
  author = {{Barret}, D. and others},
  title = "{The Athena X-ray Integral Field Unit: a consolidated design for the system requirement review of the preliminary definition phase}",
  journal = {Experimental Astronomy},
  year = 2023,
  month = apr,
  volume = {55},
  number = {2},
  pages = {373--426},
  doi = {10.1007/s10686-022-09880-7},
  archivePrefix = {arXiv},
  eprint = {2208.14562},
  primaryClass = {astro-ph.IM},
  url = {https://ui.adsabs.harvard.edu/abs/2023ExA....55..373B/abstract}
}

@ARTICLE{antonucci1993,
   author = {{Antonucci}, R.},
    title = "{Unified models for active galactic nuclei and quasars}",
  journal = {\araa},
     year = 1993,
   volume = 31,
    pages = {473-521},
      doi = {10.1146/annurev.aa.31.090193.002353},
   adsurl = {http://adsabs.harvard.edu/abs/1993ARA%26A..31..473A}
}

@ARTICLE{urry1995,
   author = {{Urry}, C.~M. and {Padovani}, P.},
    title = "{Unified Schemes for Radio-Loud Active Galactic Nuclei}",
  journal = {\pasp},
   eprint = {astro-ph/9506063},
     year = 1995,
    month = sep,
   volume = 107,
    pages = {803},
      doi = {10.1086/133630},
   adsurl = {http://adsabs.harvard.edu/abs/1995PASP..107..803U}
}

@ARTICLE{ikeda2009,
   author = {{Ikeda}, S. and {Awaki}, H. and {Terashima}, Y.},
    title = "{Study on X-Ray Spectra of Obscured Active Galactic Nuclei Based on Monte Carlo Simulation-An Interpretation of Observed Wide-Band Spectra}",
  journal = {\apj},
archivePrefix = "arXiv",
   eprint = {0810.3950},
     year = 2009,
    month = feb,
   volume = 692,
    pages = {608-617},
      doi = {10.1088/0004-637X/692/1/608},
   adsurl = {http://adsabs.harvard.edu/abs/2009ApJ...692..608I}
}

@ARTICLE{brightman2011,
   author = {{Brightman}, M. and {Nandra}, K.},
    title = "{An XMM-Newton spectral survey of 12 {$\mu$}m selected galaxies - I. X-ray data}",
  journal = {\mnras},
archivePrefix = "arXiv",
   eprint = {1012.3345},
 primaryClass = "astro-ph.HE",
     year = 2011,
    month = may,
   volume = 413,
    pages = {1206-1235},
      doi = {10.1111/j.1365-2966.2011.18207.x},
   adsurl = {http://adsabs.harvard.edu/abs/2011MNRAS.413.1206B}
}

@ARTICLE{liu2014,
   author = {{Liu}, Y. and {Li}, X.},
    title = "{An X-Ray Spectral Model for Clumpy Tori in Active Galactic Nuclei}",
  journal = {\apj},
archivePrefix = "arXiv",
   eprint = {1405.0687},
 primaryClass = "astro-ph.HE",
     year = 2014,
    month = may,
   volume = 787,
      eid = {52},
    pages = {52},
      doi = {10.1088/0004-637X/787/1/52},
   adsurl = {http://adsabs.harvard.edu/abs/2014ApJ...787...52L}
}

@ARTICLE{balokovic2018,
  author = {{Balokovi{\'c}}, M. and {Brightman}, M. and {Harrison}, F.~A. and
            {Comastri}, A. and {Ricci}, C. and others},
  title = "{New Spectral Model for Constraining Torus Covering Factors from Broadband X-Ray Spectra of Active Galactic Nuclei}",
  journal = {\apj},
  year = 2018,
  month = feb,
  volume = 854,
  eid = {42},
  pages = {42},
  doi = {10.3847/1538-4357/aaa7eb},
  archivePrefix = {arXiv},
  eprint = {1801.04938},
  primaryClass = {astro-ph.HE},
  url = {https://ui.adsabs.harvard.edu/abs/2018ApJ...854...42B/abstract}
}

@ARTICLE{buchner2019,
       author = {{Buchner}, Johannes and {Brightman}, Murray and {Nandra}, Kirpal and {Nikutta}, Robert and {Bauer}, Franz E.},
        title = "{X-ray spectral and eclipsing model of the clumpy obscurer in active galactic nuclei}",
      journal = {\aap},
         year = 2019,
        month = sep,
       volume = {629},
          eid = {A16},
        pages = {A16},
          doi = {10.1051/0004-6361/201834771},
archivePrefix = {arXiv},
       eprint = {1907.13137},
 primaryClass = {astro-ph.HE},
       adsurl = {https://ui.adsabs.harvard.edu/abs/2019A&A...629A..16B}
}

@ARTICLE{vandermeulen2024,
       author = {{Vander Meulen}, Bert and {Camps}, Peter and {Tsujimoto}, Masahiro and {Wada}, Keiichi},
        title = "{Intrinsic line profiles for X-ray fluorescent lines in SKIRT}",
      journal = {\aap},
         year = 2024,
        month = aug,
       volume = {688},
          eid = {L33},
        pages = {L33},
          doi = {10.1051/0004-6361/202451370},
archivePrefix = {arXiv},
       eprint = {2408.03367},
 primaryClass = {astro-ph.HE},
       adsurl = {https://ui.adsabs.harvard.edu/abs/2024A&A...688L..33V}
}

@ARTICLE{Furui_2016,
  author = {{Furui}, Shun'ya and {Fukazawa}, Yasushi and {Odaka}, Hirokazu and
            {Kawaguchi}, Toshihiro and {Ohno}, Masanori and others},
  title = "{X-Ray Spectral Model of Reprocess by Smooth and Clumpy Molecular Tori in Active Galactic Nuclei with the Framework MONACO}",
  journal = {\apj},
  year = 2016,
  month = feb,
  volume = {818},
  number = {2},
  eid = {164},
  pages = {164},
  doi = {10.3847/0004-637X/818/2/164},
  archivePrefix = {arXiv},
  eprint = {1602.00806},
  primaryClass = {astro-ph.HE},
  url = {https://ui.adsabs.harvard.edu/abs/2016ApJ...818..164F/abstract}
}

@ARTICLE{murphy2009,
   author = {{Murphy}, K.~D. and {Yaqoob}, T.},
    title = "{An X-ray spectral model for Compton-thick toroidal reprocessors}",
  journal = {\mnras},
archivePrefix = "arXiv",
   eprint = {0905.3188},
 primaryClass = "astro-ph.HE",
     year = 2009,
    month = aug,
   volume = 397,
    pages = {1549-1562},
      doi = {10.1111/j.1365-2966.2009.15025.x},
   adsurl = {http://adsabs.harvard.edu/abs/2009MNRAS.397.1549M}
}

@ARTICLE{yaqoob2012,
   author = {{Yaqoob}, T.},
    title = "{The nature of the Compton-thick X-ray reprocessor in NGC 4945}",
  journal = {\mnras},
archivePrefix = "arXiv",
   eprint = {1204.4196},
 primaryClass = "astro-ph.HE",
     year = 2012,
    month = jul,
   volume = 423,
    pages = {3360-3396},
      doi = {10.1111/j.1365-2966.2012.21129.x},
   adsurl = {http://adsabs.harvard.edu/abs/2012MNRAS.423.3360Y}
}

@ARTICLE{tzanavaris2019,
       author = {{Tzanavaris}, P. and {Yaqoob}, T. and {LaMassa}, S. and {Yukita}, M. and {Ptak}, A.},
        title = "{Broadband X-Ray Constraints on the Accreting Black Hole in Quasar 4C 74.26}",
      journal = {\apj},
         year = 2019,
        month = nov,
       volume = {885},
       number = {1},
          eid = {62},
        pages = {62},
          doi = {10.3847/1538-4357/ab4282},
archivePrefix = {arXiv},
       eprint = {1909.05861},
 primaryClass = {astro-ph.HE},
       adsurl = {https://ui.adsabs.harvard.edu/abs/2019ApJ...885...62T}
}

@ARTICLE{2019ApJ...885...62T,
       author = {{Tzanavaris}, P. and {Yaqoob}, T. and {LaMassa}, S. and {Yukita}, M. and {Ptak}, A.},
        title = "{Broadband X-Ray Constraints on the Accreting Black Hole in Quasar 4C 74.26}",
      journal = {\apj},
         year = 2019,
        month = nov,
       volume = {885},
       number = {1},
          eid = {62},
        pages = {62},
          doi = {10.3847/1538-4357/ab4282},
archivePrefix = {arXiv},
       eprint = {1909.05861},
 primaryClass = {astro-ph.HE},
       adsurl = {https://ui.adsabs.harvard.edu/abs/2019ApJ...885...62T}
}

@INPROCEEDINGS{Arnaud_1996,
       author = {{Arnaud}, K.~A.},
        title = "{XSPEC: The First Ten Years}",
    booktitle = {Astronomical Data Analysis Software and Systems V},
         year = 1996,
       editor = {{Jacoby}, George H. and {Barnes}, Jeannette},
       series = {Astronomical Society of the Pacific Conference Series},
       volume = {101},
        month = jan,
        pages = {17},
       adsurl = {https://ui.adsabs.harvard.edu/abs/1996ASPC..101...17A}
}

@ARTICLE{buchner2014,
  author = {{Buchner}, J. and {Georgakakis}, A. and {Nandra}, K. and {Hsu}, L. and
            {Rangel}, C. and others},
  title = "{X-ray spectral modelling of the AGN obscuring region in the CDFS: Bayesian model selection and catalogue}",
  journal = {\aap},
  year = 2014,
  month = apr,
  volume = {564},
  eid = {A125},
  pages = {A125},
  doi = {10.1051/0004-6361/201322971},
  archivePrefix = {arXiv},
  eprint = {1402.0004},
  primaryClass = {astro-ph.HE},
  url = {https://ui.adsabs.harvard.edu/abs/2014A&A...564A.125B/abstract}
}

@Book{stodden2014,
  author = 	 {{Stodden}, V. and {Leisch}, F. and {Peng}, R.~D.},
  ALTeditor = 	 {},
  title = 	     {Implementing Reproducible Research},
  publisher = 	 {CRC Press},
  year = 	 {2014},
  OPTkey = 	 {},
  OPTvolume = 	 {},
  OPTnumber = 	 {},
  OPTseries = 	 {},
  OPTaddress = 	 {},
  OPTedition = 	 {},
  OPTmonth = 	 {},
  OPTnote = 	 {},
  OPTannote = 	 {}
}

@ARTICLE{Barret_2024,
       author = {{Barret}, Didier and {Dupourqu{\'e}}, Simon},
        title = "{Simulation-based inference with neural posterior estimation applied to X-ray spectral fitting. Demonstration of working principles down to the Poisson regime}",
      journal = {\aap},
         year = 2024,
        month = jun,
       volume = {686},
          eid = {A133},
        pages = {A133},
          doi = {10.1051/0004-6361/202449214},
archivePrefix = {arXiv},
       eprint = {2401.06061},
 primaryClass = {astro-ph.IM},
       adsurl = {https://ui.adsabs.harvard.edu/abs/2024A&A...686A.133B}
}

@Book{acquaviva2023,
  author = 	 {{Acquaviva}, V.},
  ALTeditor = 	 {},
  title = 	 {Machine Learning for Physics and Astronomy},
  publisher = 	 {Princeton University Press},
  year = 	 {2023},
  OPTkey = 	 {},
  OPTvolume = 	 {},
  OPTnumber = 	 {},
  OPTseries = 	 {},
  OPTaddress = 	 {},
  OPTedition = 	 {},
  OPTmonth = 	 {},
  OPTnote = 	 {},
  OPTannote = 	 {}
}

@INPROCEEDINGS{webb2025,
       author = {{Webb}, Sara A. and {Goode}, Simon R.},
        title = "{An Astronomers Guide to Machine Learning}",
    booktitle = {IAU Symposium},
         year = 2025,
       editor = {{McIver}, J. and {Mahabal}, A. and {Fluke}, C.},
       series = {IAU Symposium},
       volume = {19},
        month = aug,
        pages = {40-49},
          doi = {10.1017/S1743921323001710},
archivePrefix = {arXiv},
       eprint = {2304.00512},
 primaryClass = {astro-ph.IM},
       adsurl = {https://ui.adsabs.harvard.edu/abs/2025IAUS..368...40W}
}

@ARTICLE{silver2023,
  author = {{Silver}, R. and {Torres-Alb{\`a}}, N. and {Zhao}, X. and {Marchesi}, S. and
            {Pizzetti}, A. and others},
  title = "{A machine learning algorithm for reliably predicting active galactic nucleus absorbing column densities}",
  journal = {\aap},
  year = 2023,
  month = jul,
  volume = {675},
  eid = {A65},
  pages = {A65},
  doi = {10.1051/0004-6361/202345980},
  archivePrefix = {arXiv},
  eprint = {2301.09598},
  primaryClass = {astro-ph.GA},
  url = {https://ui.adsabs.harvard.edu/abs/2023A&A...675A..65S/abstract}
}

@ARTICLE{dupourque2025,
       author = {{Dupourqu{\'e}}, Simon and {Barret}, Didier},
        title = "{Simulation-based inference with neural posterior estimation applied to X-ray spectral fitting: II. High-resolution spectroscopy with Athena X-IFU}",
      journal = {\aap},
         year = 2025,
        month = jul,
       volume = {699},
          eid = {A179},
        pages = {A179},
          doi = {10.1051/0004-6361/202555215},
archivePrefix = {arXiv},
       eprint = {2506.05911},
 primaryClass = {astro-ph.IM},
       adsurl = {https://ui.adsabs.harvard.edu/abs/2025A&A...699A.179D}
}

@ARTICLE{Ichinohe_2018,
       author = {{Ichinohe}, Y. and {Yamada}, S. and {Miyazaki}, N. and {Saito}, S.},
        title = "{Neural network-based preprocessing to estimate the parameters of the X-ray emission of a single-temperature thermal plasma}",
      journal = {\mnras},
         year = 2018,
        month = apr,
       volume = {475},
       number = {4},
        pages = {4739-4744},
          doi = {10.1093/mnras/sty161},
archivePrefix = {arXiv},
       eprint = {1801.06015},
 primaryClass = {astro-ph.IM},
       adsurl = {https://ui.adsabs.harvard.edu/abs/2018MNRAS.475.4739I}
}

@article{Parker_2022,
    author = {Parker, M L and Lieu, M and Matzeu, G A},
    title = {AGN X-ray spectroscopy with neural networks},
    journal = {Monthly Notices of the Royal Astronomical Society},
    volume = {514},
    number = {3},
    pages = {4061-4068},
    year = {2022},
    month = {06},
    issn = {0035-8711},
    doi = {10.1093/mnras/stac1639},
    url = {https://doi.org/10.1093/mnras/stac1639},
    eprint = {https://academic.oup.com/mnras/article-pdf/514/3/4061/44280940/stac1639.pdf},
}

@INPROCEEDINGS{gendreau2012,
       author = {{Gendreau}, Keith C. and {Arzoumanian}, Zaven and {Okajima}, Takashi},
        title = "{The Neutron star Interior Composition ExploreR (NICER): an Explorer mission of opportunity for soft x-ray timing spectroscopy}",
    booktitle = {Space Telescopes and Instrumentation 2012: Ultraviolet to Gamma Ray},
         year = 2012,
       editor = {{Takahashi}, Tadayuki and {Murray}, Stephen S. and {den Herder}, Jan-Willem A.},
       series = {Society of Photo-Optical Instrumentation Engineers (SPIE) Conference Series},
       volume = {8443},
        month = sep,
          eid = {844313},
        pages = {844313},
          doi = {10.1117/12.926396},
       adsurl = {https://ui.adsabs.harvard.edu/abs/2012SPIE.8443E..13G}
}

@ARTICLE{Tregidga_2024,
       author = {{Tregidga}, Ethan and {Steiner}, James F. and {Garraffo}, Cecilia and {Rhea}, Carter and {Aubin}, Mayeul},
        title = "{Rapid spectral parameter prediction for black hole X-ray binaries using physicalized autoencoders}",
      journal = {\mnras},
         year = 2024,
        month = apr,
       volume = {529},
       number = {2},
        pages = {1654-1666},
          doi = {10.1093/mnras/stae629},
archivePrefix = {arXiv},
       eprint = {2310.17249},
 primaryClass = {astro-ph.IM},
       adsurl = {https://ui.adsabs.harvard.edu/abs/2024MNRAS.529.1654T}
}

@ARTICLE{Dauser_2016,
  author = {{Dauser}, T. and {Garc{\'\i}a}, J. and {Walton}, D.~J. and {Eikmann}, W. and
            {Kallman}, T. and others},
  title = "{Normalizing a relativistic model of X-ray reflection. Definition of the reflection fraction and its implementation in relxill}",
  journal = {\aap},
  year = 2016,
  month = may,
  volume = {590},
  eid = {A76},
  pages = {A76},
  doi = {10.1051/0004-6361/201628135},
  archivePrefix = {arXiv},
  eprint = {1601.03771},
  primaryClass = {astro-ph.HE},
  url = {https://ui.adsabs.harvard.edu/abs/2016A&A...590A..76D/abstract}
}

@ARTICLE{Zhang2025,
  author = {{Zhang}, Rui and {Guo}, Xiaotong and {Gu}, Qiusheng and {Fang}, Guanwen and
            {Xu}, Jun and others},
  title = "{Identifying Compton-thick Active Galactic Nuclei with a Machine Learning Algorithm in Chandra Deep Field-South}",
  journal = {\apj},
  year = 2025,
  month = jul,
  volume = {987},
  number = {1},
  eid = {46},
  pages = {46},
  doi = {10.3847/1538-4357/addaab},
  archivePrefix = {arXiv},
  eprint = {2505.21105},
  primaryClass = {astro-ph.GA},
  url = {https://ui.adsabs.harvard.edu/abs/2025ApJ...987...46Z/abstract}
}

@ARTICLE{yaqoob2011,
   author = {{Yaqoob}, T. and {Murphy}, K.~D.},
    title = "{The Compton shoulder of the Fe K{$\alpha$} fluorescent emission line in active galactic nuclei}",
  journal = {\mnras},
archivePrefix = "arXiv",
   eprint = {1010.5262},
 primaryClass = "astro-ph.HE",
     year = 2011,
    month = mar,
   volume = 412,
    pages = {277-286},
      doi = {10.1111/j.1365-2966.2010.17902.x},
   adsurl = {http://adsabs.harvard.edu/abs/2011MNRAS.412..277Y}
}

@ARTICLE{lamassa2014,
  author = {{LaMassa}, S.~M. and {Yaqoob}, T. and {Ptak}, A.~F. and {Jia}, J. and
            {Heckman}, T.~M. and others},
  title = "{Delving Into X-ray Obscuration of Type 2 AGN, Near and Far}",
  journal = {\apj},
  year = 2014,
  month = may,
  volume = 787,
  eid = {61},
  pages = {61},
  doi = {10.1088/0004-637X/787/1/61},
  archivePrefix = {arXiv},
  eprint = {1404.0012},
  primaryClass = {astro-ph.HE},
  url = {https://ui.adsabs.harvard.edu/abs/2014ApJ...787...61L/abstract}
}

@ARTICLE{yaqoob2016,
   author = {{Yaqoob}, T. and {Turner}, T.~J. and {Tatum}, M.~M. and {Trevor}, M. and 
	{Scholtes}, A.},
    title = "{No signatures of black hole spin in the X-ray spectrum of the Seyfert 1 galaxy Fairall 9}",
  journal = {\mnras},
archivePrefix = "arXiv",
   eprint = {1607.07125},
 primaryClass = "astro-ph.HE",
     year = 2016,
    month = nov,
   volume = 462,
    pages = {4038-4054},
      doi = {10.1093/mnras/stw1824},
   adsurl = {http://adsabs.harvard.edu/abs/2016MNRAS.462.4038Y}
}

@ARTICLE{harrison2013,
  author = {{Harrison}, F.~A. and others},
  title = "{The Nuclear Spectroscopic Telescope Array (NuSTAR) High-energy X-Ray Mission}",
  journal = {\apj},
  year = 2013,
  month = jun,
  volume = 770,
  eid = {103},
  pages = {103},
  doi = {10.1088/0004-637X/770/2/103},
  archivePrefix = {arXiv},
  eprint = {1301.7307},
  primaryClass = {astro-ph.IM},
  url = {https://ui.adsabs.harvard.edu/abs/2013ApJ...770..103H/abstract}
}

@ARTICLE{torres-alba2023,
  author = {{Torres-Alb{\`a}}, N. and {Marchesi}, S. and {Zhao}, X. and {Cox}, I. and
            {Pizzetti}, A. and others},
  title = "{Hydrogen column density variability in a sample of local Compton-thin AGN}",
  journal = {\aap},
  year = 2023,
  month = oct,
  volume = {678},
  eid = {A154},
  pages = {A154},
  doi = {10.1051/0004-6361/202345947},
  archivePrefix = {arXiv},
  eprint = {2301.07138},
  primaryClass = {astro-ph.GA},
  url = {https://ui.adsabs.harvard.edu/abs/2023A&A...678A.154T/abstract}
}

@ARTICLE{pizzetti2025,
  author = {{Pizzetti}, A. and {Torres-Alb{\`a}}, N. and {Marchesi}, S. and
            {Buchner}, J. and {Cox}, I. and others},
  title = "{Hydrogen Column Density Variability in a Sample of Local Compton-thin AGN II}",
  journal = {\apj},
  year = 2025,
  month = feb,
  volume = {979},
  number = {2},
  eid = {170},
  pages = {170},
  doi = {10.3847/1538-4357/ad9c64},
  archivePrefix = {arXiv},
  eprint = {2403.06919},
  primaryClass = {astro-ph.HE},
  url = {https://ui.adsabs.harvard.edu/abs/2025ApJ...979..170P/abstract}
}

@ARTICLE{jana2022,
  author = {{Jana}, Arghajit and {Ricci}, Claudio and {Naik}, Sachindra and
            {Tanimoto}, Atsushi and {Kumari}, Neeraj and others},
  title = "{Absorption variability of the highly obscured active galactic nucleus NGC 4507}",
  journal = {\mnras},
  year = 2022,
  month = jun,
  volume = {512},
  number = {4},
  pages = {5942--5959},
  doi = {10.1093/mnras/stac799},
  archivePrefix = {arXiv},
  eprint = {2203.10550},
  primaryClass = {astro-ph.HE},
  url = {https://ui.adsabs.harvard.edu/abs/2022MNRAS.512.5942J/abstract}
}

@ARTICLE{6773024,
  author={Shannon, C. E.},
  journal={The Bell System Technical Journal}, 
  title={A mathematical theory of communication}, 
  year={1948},
  volume={27},
  number={3},
  pages={379-423},
  doi={10.1002/j.1538-7305.1948.tb01338.x}}

@ARTICLE{Rosales-Ortega_2012,
       author = {{Rosales-Ortega}, F.~F. and {Arribas}, S. and {Colina}, L.},
        title = "{Integrated spectra extraction based on signal-to-noise optimization using integral field spectroscopy}",
      journal = {\aap},
         year = 2012,
        month = mar,
       volume = {539},
          eid = {A73},
        pages = {A73},
          doi = {10.1051/0004-6361/201117774},
archivePrefix = {arXiv},
       eprint = {1112.5357},
 primaryClass = {astro-ph.IM},
       adsurl = {https://ui.adsabs.harvard.edu/abs/2012A&A...539A..73R}
}

@ARTICLE{Torres-Alba_2023,
  author = {{Torres-Alb{\`a}}, N. and {Marchesi}, S. and {Zhao}, X. and {Cox}, I. and
            {Pizzetti}, A. and others},
  title = "{Hydrogen column density variability in a sample of local Compton-thin AGN}",
  journal = {\aap},
  year = 2023,
  month = oct,
  volume = {678},
  eid = {A154},
  pages = {A154},
  doi = {10.1051/0004-6361/202345947},
  archivePrefix = {arXiv},
  eprint = {2301.07138},
  primaryClass = {astro-ph.GA},
  url = {https://ui.adsabs.harvard.edu/abs/2023A&A...678A.154T/abstract}
}

@ARTICLE{Germain_2015,
       author = {{Germain}, Mathieu and {Gregor}, Karol and {Murray}, Iain and {Larochelle}, Hugo},
        title = "{MADE: Masked Autoencoder for Distribution Estimation}",
      journal = {arXiv e-prints},
         year = 2015,
        month = feb,
          eid = {arXiv:1502.03509},
        pages = {arXiv:1502.03509},
          doi = {10.48550/arXiv.1502.03509},
archivePrefix = {arXiv},
       eprint = {1502.03509},
 primaryClass = {cs.LG},
       adsurl = {https://ui.adsabs.harvard.edu/abs/2015arXiv150203509G}
}

@ARTICLE{Gloeckler_2024,
       author = {{Gloeckler}, Manuel and {Deistler}, Michael and {Weilbach}, Christian and {Wood}, Frank and {Macke}, Jakob H.},
        title = "{All-in-one simulation-based inference}",
      journal = {arXiv e-prints},
         year = 2024,
        month = apr,
          eid = {arXiv:2404.09636},
        pages = {arXiv:2404.09636},
          doi = {10.48550/arXiv.2404.09636},
archivePrefix = {arXiv},
       eprint = {2404.09636},
 primaryClass = {cs.LG},
       adsurl = {https://ui.adsabs.harvard.edu/abs/2024arXiv240409636G}
}

@ARTICLE{Barret_3_2026,
       author = {{Barret}, Didier and {Dupourqu{\'e}}, Simon},
        title = "{Simulation-based inference with neural posterior estimation applied to X-ray spectral fitting: III. Deriving exact posteriors with dimension reduction and importance sampling}",
      journal = {\aap},
         year = 2026,
        month = apr,
       volume = {708},
          eid = {A280},
        pages = {A280},
          doi = {10.1051/0004-6361/202557639},
archivePrefix = {arXiv},
       eprint = {2512.16709},
 primaryClass = {astro-ph.IM},
       adsurl = {https://ui.adsabs.harvard.edu/abs/2026A&A...708A.280B}
}

@misc{mlmytorus_github,
  author       = {Daza-Perilla, Ingrid Vanessa and Tzanavaris, Panayiotis and Madurga-Favieres, V. and Yukita, M. and Ptak, A. and Yaqoob, T.},
  title        = {{ML\_MYTORUS}: Neural Posterior Estimation software and trained model},
  howpublished = {\url{https://github.com/vanedaza/viva-streamlit}},
  year         = {2026}
}

@misc{mlmytorus_webapp,
  author       = {Daza-Perilla, Ingrid Vanessa and Tzanavaris, Panayiotis and Madurga-Favieres, V. and Yukita, M. and Ptak, A. and Yaqoob, T.},
  title        = {{ML\_MYTORUS} Web Interface},
  howpublished = {\url{https://web-ml-mytorus.streamlit.app/}},
  year         = {2026}
}
\clearpage


\appendix

\section{Spectra and Univariate analysis}\label{app:spectra_univariate}
In order to evaluate the statistical properties of the simulated spectra and their relation to the performance in the determination of the physical parameters with ML, we first analyzed the distributions of the S/N obtained from both the observation based and uniform grids (see Figure~\ref{fig:hist_sn}). These histograms highlight the different regimes sampled by the two strategies, with the observational grid reproducing the predominance of low S/N spectra found in real \nustar\ data, while the uniform grid yields a broader distribution extending toward higher values. In parallel, we examined how additional observational quantities, exposure time, \nhgal, and $z$, correlate with the four main physical parameters through IM (see Figure~\ref{fig:IM_grid}). These results highlight that observation conditions carry non-negligible information about the physical parameter space, supporting their inclusion as auxiliary inputs in the ML Algorithm.

\begin{figure}[!htbp]
    \centering
    \includegraphics[width=0.85\linewidth]{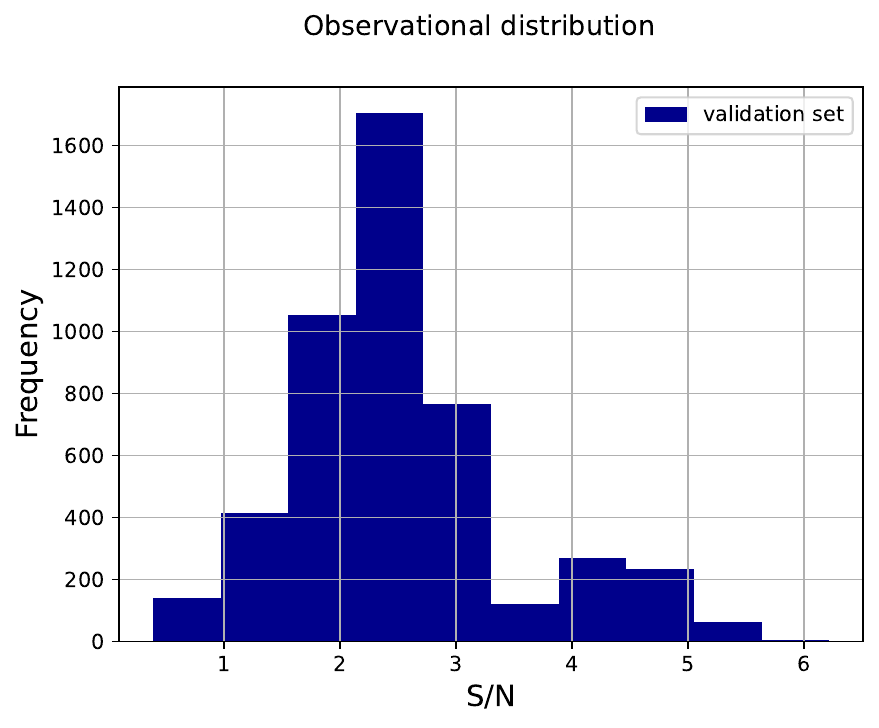}
    \includegraphics[width=0.85\linewidth]{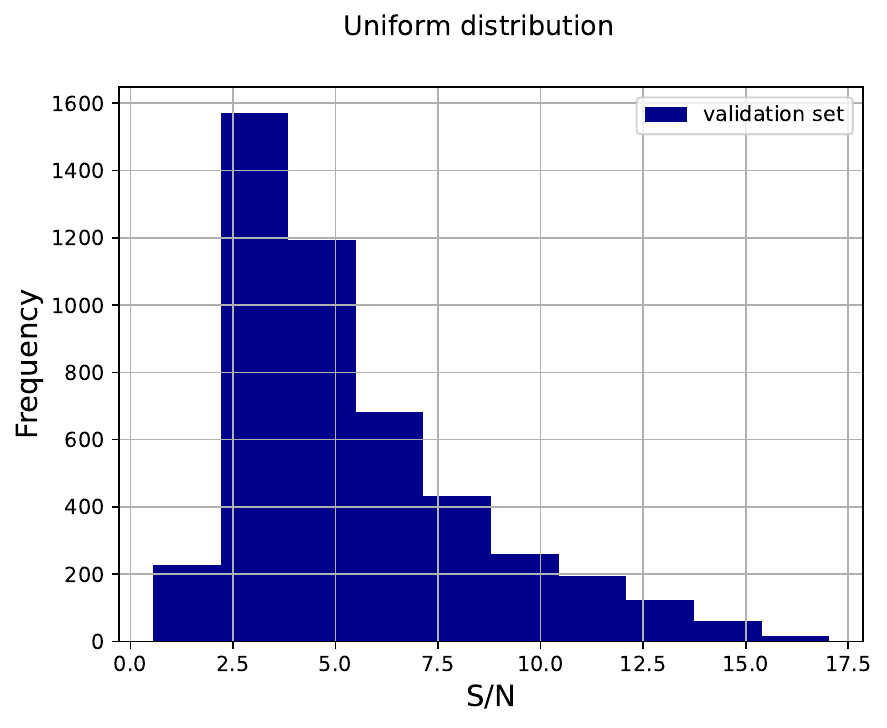}
    \caption{Histograms of S/N distributions for simulated spectra. {\it Upper panel}: observational-based grid, which reproduces the predominance of low S/N values found in observed \nustar\ spectra, with a narrow distribution centered around S/N \sss2.5. {\it Lower panel}: uniform grid, which produces a broader distribution shifted toward higher S/N values (mean \sss 5.4), reflecting conditions not typically observed.}
    \label{fig:hist_sn}
\end{figure}
\begin{figure}
    \centering
    \includegraphics[width=1\linewidth]{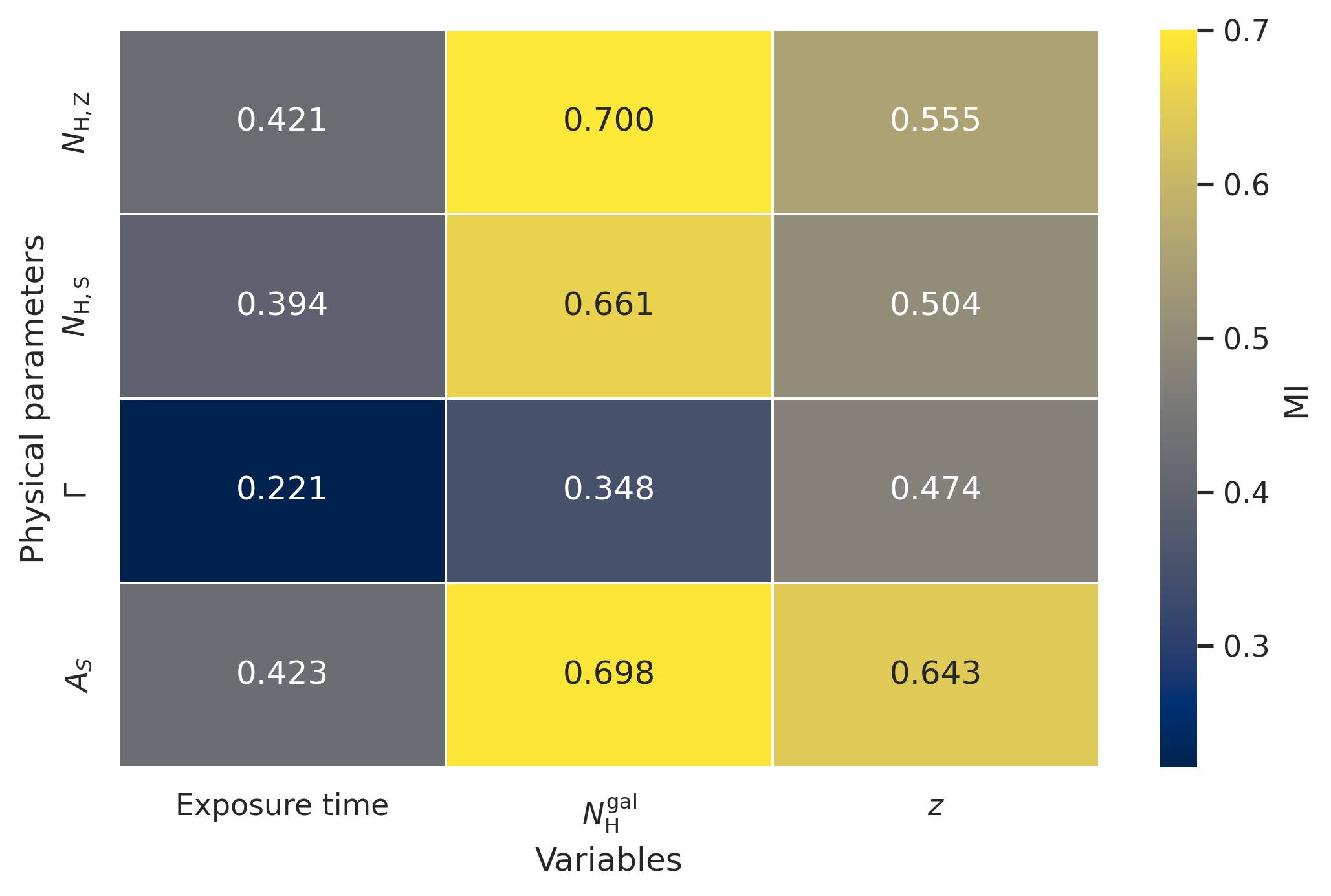}
    
    \label{fig:IM_grid}
    \caption{MI heatmap between, on the one had, the four \myt\ physical parameters and, on the other hand, exposure time, \nhgal, and $z$. Each grid cell quantifies the level of statistical dependence.}
\end{figure}

\section{Workflow and Architecture of SBI-NPE}\label{app:C}
\subsection{training}
The simulation-based inference NPE framework is trained to recover the physical parameters 
$\bm{\theta} = [\theta_1, \theta_2, \theta_3, \theta_4]$ 
from an input consisting of a spectrum $\bm{x}$, together with $z$, \nhgal, and exposure time, that is  
$x = [x_1, x_2, \dots, x_{679}, z,$ \nhgal$, \rm{exp.\ time}]$.

During training, the model learns an invertible transformation between $\bm{\theta}$ and a latent variable $\bm{z} \sim \mathcal{N}(0,I)$ through a sequence of five autoregressive flows. As illustrated in Figure~\ref{fig:nn}, the training process begins with a component-wise affine transformation (PointwiseAffine) that rescales and shifts each parameter as $\theta'_i = a_i \theta_i + b_i$, followed by a series of MADE layers. Each MADE processes the reparameterized parameters through masked linear layers with ReLU (Rectified Linear Unit) activation functions, defined as $\text{ReLU}(\bm{x}) = \max(0, \bm{x})$. The masked linear layers enforce the autoregressive structure by restricting dependencies so that each parameter $\theta_i$ depends only on previous parameters $\theta_{<i}$, while the ReLU activations introduce the non-linearity needed to learn complex relationships conditioned on the observed spectrum. The spectrum information is incorporated through a context layer that transforms $\bm{x}$ from 769 to 50 dimensions and is added to the main processing path.  

Each MADE outputs scale and shift parameters $[s_i, \mu_i]$ that define the inverse affine transformation (Eq.~\ref{eq:affine_transform}) 
$f^{-1} = z_i = \frac{\theta'_i - \mu_i}{s_i}$,  
with the Jacobian determinant computed as $\log \det J = \sum \log |s_i|$, required for the evaluation of the $\mathcal{L}_{\rm{NLL}}$ (see Eq.~\ref{eq:affine_transform}). The outputs of the MADEs are randomly permuted to increase mixing and ensure that all parameters are effectively transformed.  
The final latent vector $\bm{z}$ follows a standard normal distribution $\mathcal{N}(0,I)$, allowing the exact log-probability computation through the change-of-variables theorem: $\log p_\theta(\bm{\theta}|\bm{x})$ (see Eq.\ref{eq:2})
which is then used in the minimization of $\mathcal{L}_{\rm{NLL}}$, encouraging the model to assign high probability to the true simulated parameters.  

\begin{figure*}
    \centering
    \includegraphics[width=1\linewidth]{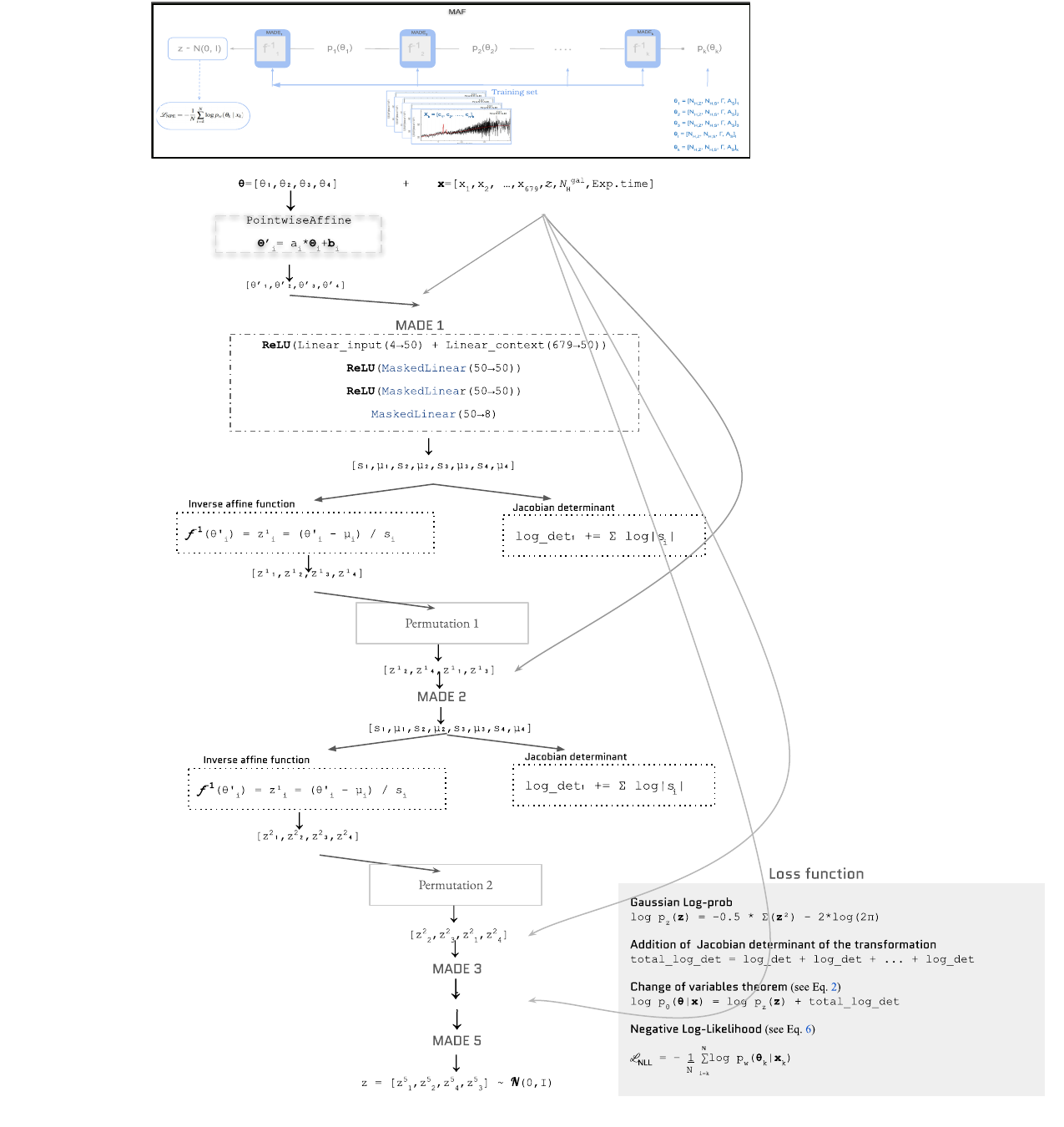}
    \caption{
    Training architecture of the NPE framework. 
    The upper panel shows the global workflow where the physical parameters $\bm{\theta}$ and the observed spectrum $\bm{x}$ are processed through a normalizing flow to produce a latent representation $\bm{z} \sim \mathcal{N}(0,I)$.  
    The lower panel details the sequential operations within each MADE layer:  
    (1) input parameters undergo a component-wise affine transformation,  
    (2) the spectral context is incorporated through a linear transformation that conditions each MADE layer,  
    (3) masked linear layers with ReLU activations enforce autoregressive dependencies,  
    (4) the output provides scale $s_i$ and shift $\mu_i$ parameters for the inverse transformation $z_i = (\theta'_i - \mu_i)/s_i$,  
    (5) Jacobian determinants are accumulated for likelihood computation, and  
    (6) permutations between layers ensure effective mixing.  
    The final loss function combines the Gaussian log-probability of the latent variables with the accumulated Jacobian determinants via the change-of-variables theorem, training the network to maximize the likelihood of the observed parameter–spectrum pairs (minimization of $\mathcal{L}_{\rm{NLL}}$).}
\label{fig:nn}
\end{figure*}

\subsection{Validation and Testing}

Once the NPE framework is trained, validation and testing are performed by reversing the normalizing flow to generate posterior distributions for the physical parameters $\bm{\theta} = [\theta_1, \theta_2, \theta_3, \theta_4]$ given an observed spectrum $\bm{x} = [x_1, x_2, \dots, x_{679}, z, N_H^{\text{gal}}, \text{Exp.time}]$. As illustrated in Figure~\ref{fig:sampling}, the inference process begins by sampling multiple latent vectors $\bm{z} \sim \mathcal{N}(0,I)$ (1000 samples) from the standard normal distribution. Each sampled latent vector is then transformed through the inverse normalizing flow, starting with MADE 5 and proceeding backwards through the architecture. 

At each MADE layer, the sampled latent variables are processed through the same masked linear layers with ReLU activations used during training. During inference, the observed spectrum $\bm{x}$ is fed into the MADE neural network (with parameters fixed after training), and the network dynamically computes the scale and shift parameters $[s_i, \mu_i]$ specific to that spectrum by processing the information through its connected layers. These values are the outputs of the neural network when evaluating the particular features of the observed spectrum, not fixed parameters learned in training. The forward affine transformation $f(\bm{z}) = \theta^{(l)}_i = z_i \cdot s_i + \mu_i$ is applied at each layer to progressively transform the latent samples toward the parameter space. 

Between layers, the same permutation patterns used during training are applied in reverse order to ensure adequate mixing of variables. After passing through the five MADE layers and their corresponding inverse permutations, the final component-wise affine transformation $\theta_i = (\theta'_i - b_i)/a_i$ recovers the physical parameters on their original scale. This process generates 1000 samples from the posterior distribution $p(\bm{\theta}|\bm{x})$ for each parameter, allowing for a thorough quantification of uncertainty through the full posterior distribution rather than point estimates. The mode of each marginal distribution serves as the predicted parameter value, while the spread quantifies the associated uncertainty using robust metrics such as $\sigma_{68} = (P[84] - P[16])/2$, where $P[x]$ denotes the $x$-th percentile of the distribution.

\begin{figure*}
    \centering
    \includegraphics[width=1\linewidth]{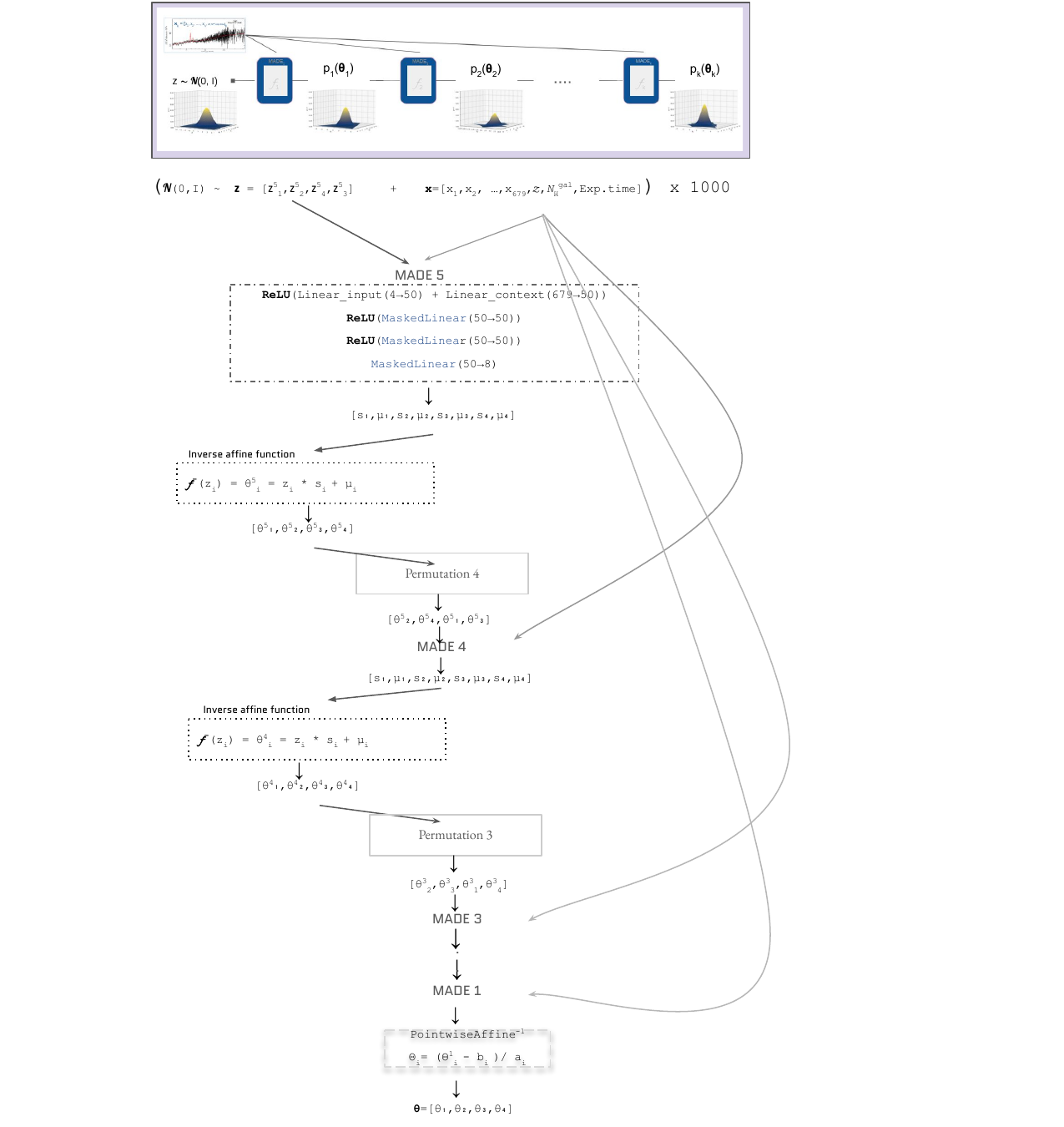}
    \caption{Inference architecture of the NPE framework for posterior sampling. The upper panel shows the global workflow where latent samples $\bm{z} \sim \mathcal{N}(0,I)$ are transformed through the inverse normalizing flow conditioned on an observed spectrum $\bm{x}$ to generate posterior samples of physical parameters $\bm{\theta}$. The lower panel details the sequential operations during inference: (1) multiple latent vectors are sampled from the standard normal distribution, (2) each MADE layer processes the samples through masked linear layers with ReLU activations, conditioned by the observed spectrum through the same context transformation used in training, (3) forward affine transformations $\theta^{(l)}_i = z_i \cdot s_i + \mu_i$ progressively map samples toward the parameter space, (4) inverse permutations (applied in reverse order from training) ensure proper variable transformation, and (5) the final inverse pointwise affine transformation recovers parameters in their original scale. This process generates a full posterior distribution for each parameter, enabling comprehensive uncertainty quantification through distributional analysis rather than point estimates.
    }
    \label{fig:sampling}
\end{figure*}

%


\end{document}